\documentclass[journal]{IEEEtran}
\usepackage{amsmath,amsfonts,amsthm}
\usepackage{array}
\usepackage[caption=false,font=normalsize,labelfont=sf,textfont=sf]{subfig}
\usepackage{textcomp}
\usepackage{stfloats}
\usepackage{url}
\usepackage{verbatim}
\usepackage{graphicx}
\usepackage{xcolor}
\usepackage{nomencl}
\usepackage{etoolbox}
\usepackage{ifthen}
\usepackage{booktabs}
\usepackage[ruled]{algorithm2e}

\usepackage{booktabs} % For formal tables
\usepackage{amssymb}  % Provides \varnothing and other symbols

\usepackage{multirow}
\usepackage{bm}
\usepackage{amsmath}
\usepackage{makecell}
\usepackage{enumitem}

\allowdisplaybreaks 
\usepackage[subtle,tracking=normal]{savetrees}
\usepackage{cite}

\usepackage{xcolor}  % 导入颜色包

\makenomenclature
\def\BibTeX{{\rm B\kern-.05em{\sc i\kern-.025em b}\kern-.08em
    T\kern-.1667em\lower.7ex\hbox{E}\kern-.125emX}}
\usepackage{balance}

\begin{document}

\bstctlcite{IEEEexample:BSTcontrol}
\title{Full Timescale Hierarchical MPC-MTIP Framework \\for Hybrid Energy Storage Management in\\ Low-Carbon Industrial Microgrid}

\author{\IEEEauthorblockN{Daniyaer Paizulamu, \textit{Graduate Student Member, IEEE}}, \IEEEauthorblockN{Lin Cheng, \textit{Senior Member, IEEE}},\\
\IEEEauthorblockN{Ning Qi, \textit{Member, IEEE}},
\IEEEauthorblockN{Zhengmao Li, \textit{Member, IEEE}},
\IEEEauthorblockN{Nikos D. Hatziargyriou, \textit{Life Fellow, IEEE}}

%\IEEEauthorblockN{Helin Xu},\\
%\IEEEauthorblockN{Yingrui Zhuang, \textit{Student Member, IEEE}},

\thanks{%Manuscript created xxx, 2024; revised xxxx, 2024; accepted xxxx 2024.
This work was supported by Smart Grid-National Sclence and Technology Major Project (No. 2025ZD0806304).
(\textit{Corresponding author: Ning Qi.})

Daniyaer Paizulamu, Lin Cheng are with the State Key Laboratory of Power System Operation and Control, Department of Electrical Engineering, Tsinghua University, Beijing 100084, China (e-mail: dnyepzlm22@mails.tsinghua.edu.cn; chenglin@mail.tsinghua.edu.cn).

Ning Qi is with the Department of Earth and Environmental Engineering, Columbia University, New York, NY 10027, USA (e-mail: nq2176@columbia.edu). 

Zhengmao Li is with the school of Electrical Engineering, Aalto University, 
Finland (zhengmao.li@aalto.fi).

Nikos D. Hatziargyriou is with the School of Electrical and Computer Engineering, National Technical University of Athens, Athens 15773, Greece (e-mail: nh@power.ece.ntua.gr).

}
\vspace{-1.5em}
}

\markboth{IEEE TRANSACTIONS ON SUSTAINABLE ENERGY,~Vol.~X, No.~X, XX Month~2026}
{How to Use the IEEEtran \LaTeX \ Templates}

\maketitle

\IEEEaftertitletext{\vspace{-1\baselineskip}}

\begin{abstract}
Uncertainties in balancing generation and load in low-carbon industrial microgrids (IMGs) make hybrid energy storage systems (HESS) crucial for their stable and economic operation. Existing model predictive control (MPC) techniques typically enforce periodic state of charge (SOC) constraints to maintain long term stability. However, these hard constraints compromise dispatch flexibility near the end of the prediction horizon, preventing sufficient energy release during critical peaks and leading to optimization infeasibility. This paper eliminates the periodic SOC constraints of individual storage units and proposes a novel full-timescale hierarchical MPC scheduling framework. 
Specifically, comprehensive physical and cost models are established for the HESS composed of flywheel, battery, compressed-air, and hydrogen-methanol energy storage. The control problem is decoupled into a hierarchical MPC architecture.
Furthermore, a novel adaptive feedback mechanism based on micro trajectory inverse projection (MTIP) is embedded into the scheduling process, accurately mapping the high frequency dynamic buffering capabilities of lower tier storages into the upper decision space to generate dynamic boundaries. Experiments using 14 consecutive months of second-level data from a real-world IMG validate the effectiveness of the proposed method, demonstrating its significant superiority over existing approaches. By effectively preventing limit violations and deadlocks in lower-tier storages under extreme fluctuations, it achieves a 97.4\% net load smoothing rate and a 62.2\% comprehensive cycle efficiency.

\end{abstract}
\mbox{}

\vspace{-0.5cm}
\section{Introduction}\label{Introduction}

\IEEEPARstart{A}{s} power systems undergo a green transition, the industrial sector remains a primary domain of energy consumption and carbon emissions\cite{TOP1}. Developing industrial microgrids (IMGs) effectively reduces industrial carbon emissions\cite{TOP2}. However, the coupled generation-load uncertainties introduce multi-timescale unbalanced components into the net load profile\cite{TOP3}, creating technical challenges in power system operation. Hybrid energy storage systems (HESS) provide an effective way to ensure the secure, stable, and economic operation of IMG.

HESS energy management methods oriented toward net load smoothing are crucial for enhancing IMG stability. Existing mainstream strategies primarily include the following: Filtering-based methods \cite{Filter1}, which decompose net load demand into high  and low frequency components, assigning them to high-power and high-energy-density storages, respectively \cite{Filter2}. While effectively smoothing power fluctuations and reducing computational burden, they struggle to handle strict physical boundary constraints such as equipment capacity and state of charge (SOC)\cite{Filter3}. Optimization approaches such as stochastic optimization (SO)\cite{SO}, distributionally robust optimization (DRO)\cite{DRO}, and chance-constrained optimization (CCO)\cite{CCO} can deal with scenario uncertainties and achieve global optima. However, they fail to balance economic efficiency with robustness\cite{CCO2}, and their exhaustive solving processes cannot meet the second-level real time dispatch requirements of microgrids. Neural network methods\cite{NN1},\cite{NN2}, rely heavily on historical data, lack physical interpretability, and cannot provide safety guarantees under sudden extreme conditions. To overcome these limitations, model predictive control (MPC) leverages predictive models and receding horizon optimization to achieve closed-loop real time optimization while strictly enforcing multiple physical constraints\cite{MPC1}. Notably, multi-layer MPC naturally aligns with multi-timescale and multi-variable coordinated control, providing an efficient framework for the dynamic power allocation of cross-scale heterogeneous energy storage systems\cite{MPC2}.

To guarantee capacity availability for subsequent dispatch, existing MPC-based net load smoothing methods typically enforce periodic SOC constraints on the HESS\cite{MPC3},\cite{MPC4}. However, these hard constraints compromise the dispatch flexibility near the end of the prediction horizon, preventing sufficient energy release during critical peaks\cite{MPC5}. This forces reliance on grid power or energy curtailment\cite{MPC6}, thereby deteriorating both smoothing efficacy and economic performance. More critically, under accumulated prediction errors\cite{MPC7}, initial SOC deviations, or pronounced efficiency disparities among HESS\cite{MPC8}, periodic constraints frequently lead to optimization infeasibility or severe compulsory recharging, ultimately distorting lifespan evaluations. To address this limitation, in \cite{MPC_SOC1}, periodic constraints are relaxed into soft constraints via penalty terms. By utilizing a terminal cost and region, the feasible domain is enlarged to mitigate conservatism while preserving system stability. A novel economic model predictive control (EMPC) framework is proposed in \cite{MPC_SOC2}, which discards terminal constraints and steady-state prior knowledge, replacing them with a state increment penalty. In \cite{MPC_SOC3}, a linear discount factor is applied to the stage cost, attenuating the adverse impact of terminal states within the prediction horizon on current decisions. In \cite{MPC_SOC4}, \cite{MPC_SOC5}, reference trajectories are employed to substitute periodic constraints, strictly enforcing the terminal states to exactly reach these trajectories at the end of the prediction horizon. By simultaneously optimizing both reference and control sequences online, recursive feasibility is guaranteed and adaptability to online variations is enhanced. A dynamic SOC limit-violation prevention mechanism is introduced in \cite{MPC_SOC6}, compelling the HESS to reduce power when approaching SOC limits. Although this fails to resolve the periodic SOC constraint issue, it effectively prevents prolonged storage stagnation. In \cite{MPC_SOC7}, periodic SOC constraints are eliminated, and a logic-rule-based upper-level power management algorithm is designed solely to prevent overcharging and overdischarging, neglecting the dynamic coordinated operation of the HESS.

In this paper, a hierarchical MPC scheduling framework is proposed to maximize the dispatch potential of the HESS by eliminating the periodic SOC constraints of individual storage units and introducing an MTIP-based coordination mechanism. Using real-world IMG data, the experimental results demonstrate that the proposed method effectively smooths the net load while ensuring operational efficiency and economic benefits. Specifically, the main contributions are as follows:

\begin{enumerate}
    \item \textit{Full Timescale HESS Model:} Comprehensive physical and cost equivalent models incorporating flywheel energy storage system (FESS),  dynamically reconfigurable battery network (DRBN), compressed air energy storage system (CAES), and hydrogen-methanol energy storage system (HMES) are established to satisfy the full-timescale net load smoothing demands of IMG.
    \item \textit{Hierarchical MPC Scheduling Framework:} The control domain is decoupled into a hierarchical architecture corresponding to long-term HMES scheduling, hourly CAES, minute-level DRBN, and second-level flywheel virtual inertia control. By strictly enforcing multiple physical boundaries, it efficiently coordinates the HESS response rates and capacity distribution, achieving an optimal balance between computational burden and multi-scale control performance.
    \item \textit{MTIP-based Adaptive Feedback Mechanism:} Eliminating traditional periodic SOC constraints, the proposed MTIP mechanism accurately maps the high-frequency dynamic buffering capabilities of lower-tier storages into the upper decision space to generate dynamic boundaries. This effectively prevents limit violations and deadlocks in lower-tier storages under extreme fluctuations, significantly enhancing cross-layer economic synergy while ensuring rigorous solvability.
    \item \textit{Real-world IMG Experimental Validation:} Using 14 consecutive months of second-level data from a practical IMG, the effectiveness, robustness, and economic efficiency of the proposed HMPC-MTIP approach under complex real-world conditions is extensively validated.
\end{enumerate}

The remainder of the paper is organized as follows. Section \ref{Modeling} establishes the full timescale mechanism and cost models of the HESS. Section \ref{Method} proposes the hierarchical MPC framework and the MTIP feedback mechanism for HESS. Experimental validation based on a real-world IMG is presented in Section \ref{Experiments}. Finally, conclusions are summarized in Section \ref{CONCLUSION}.

\section{Hybrid Energy Storage Components Modeling}\label{Modeling}
\begin{figure}[htbp] 
    \vspace{-0.2cm}
    \centerline{\includegraphics[width=0.9\columnwidth]{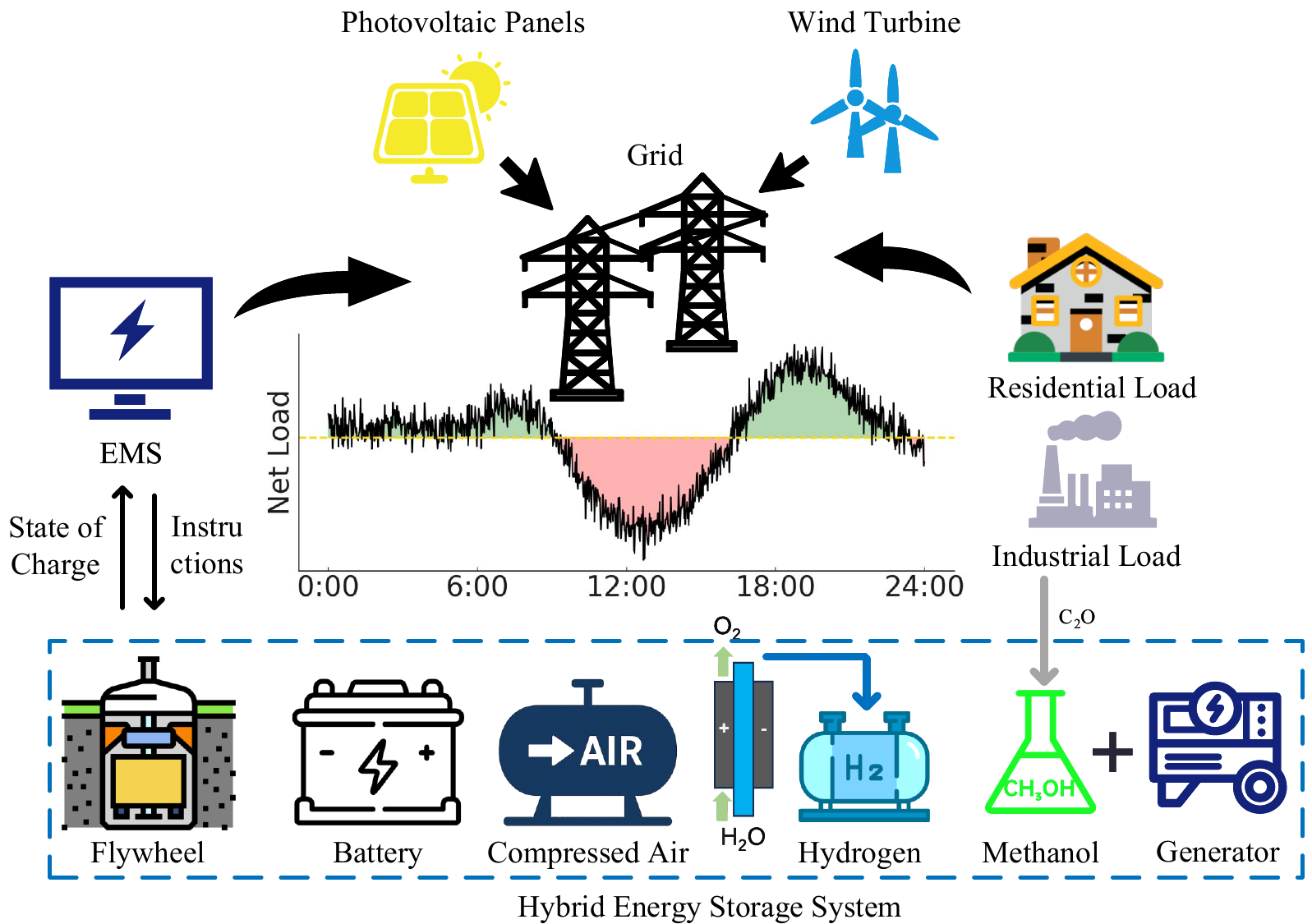}}
    \caption{IMG structure and composition of the full timescale HESS.}
    \label{shiyitu}
    \vspace{-0.6cm}
\end{figure}

This section models the five types of energy storage devices in HESS, calibrates the operational constraints in the context of IMG, and comprehensively considers various costs and full lifecycle degradation models. The expressions for the instantaneous levelized cost of electricity (LCOE) for each energy storage device are derived to support the subsequent sections. Fig.~\ref{shiyitu} illustrates the architecture of IMG and composition of the full timescale HESS.

\subsection{Flywheel Energy Storage Systems Modeling}\label{Flywheel Modeling}
\begin{figure}[htbp] 
    \vspace{-0.2cm}
    \centerline{\includegraphics[width=1\columnwidth]{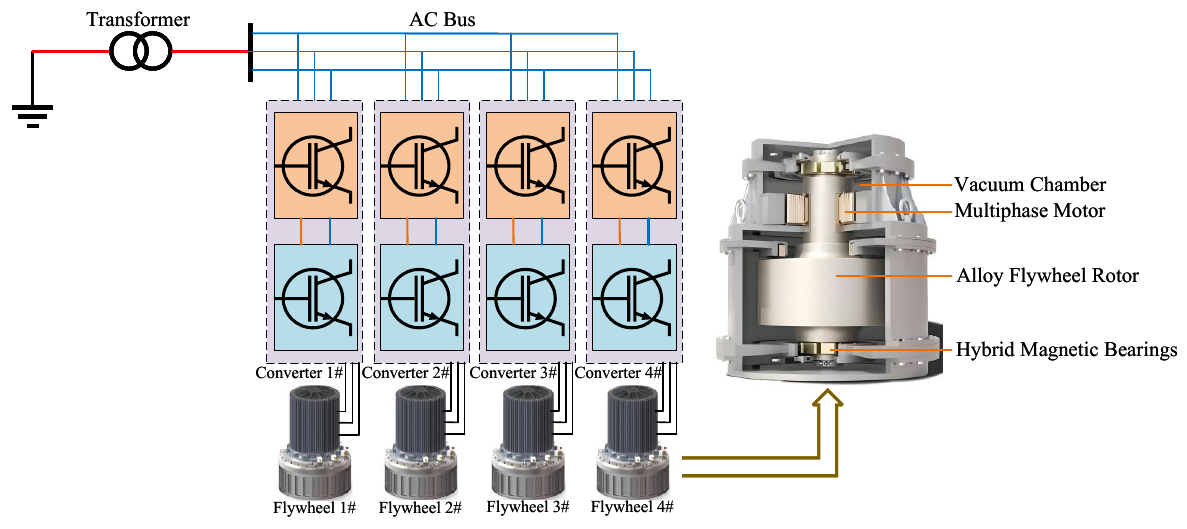}}
    \caption{Composition and topology of flywheel energy storage system.}
    \label{Fess}
    \vspace{-0.5cm}
\end{figure}

FESS provides effective inertia support and primary frequency regulation for IMG with high renewable penetration. During grid-connected operation, the FESS enhances power system inertia, supplying virtual inertia that exceeds its intrinsic mechanical inertia. As illustrated in Fig.~\ref{Fess}, a typical FESS comprises flywheel rotor within vacuum enclosure, magnetic bearings, high-speed motor-generator, and bidirectional converter. This study utilizes inertial flywheels with high power density:
\begin{subequations}\label{eq:fw_group}
\begin{align}
E^{\mathrm{F}}_t
&= \frac{1}{2}\sum\limits_{i=1}^{n} J_{\mathrm{fw,i}}\,\omega_{\mathrm{efw}}^{2}, \label{eq:fw_group_a}\\[2pt]
\Delta E^{\mathrm{F}}_t
&= \frac{1}{2} \sum\limits_{i=1}^{n} J_{\mathrm{fw,i}}
   \Big[\omega_{\mathrm{efw0}}^{2}
      -\big(\omega_{\mathrm{efw0}}-\Delta\omega_{\mathrm{efw0}}\big)^{2}\Big] \notag\\
&= \frac{1}{2} \sum\limits_{i=1}^{n} J_{\mathrm{vir,i}}
   \Big[\omega_{\mathrm{sfw}}^{2}
      -\big(\omega_{\mathrm{sfw}}-\Delta\omega_{\mathrm{sfw}}\big)^{2}\Big], \label{eq:fw_group_b}\\[2pt]
J_{\mathrm{vir}}
&= \sum\limits_{i=1}^{n} J_{\mathrm{fw,i}}\,
   \frac{\omega_{\mathrm{efw0}}}{\omega_{\mathrm{sfw0}}}\,
   \frac{\Delta\omega_{\mathrm{efw0}}}{2\pi\,\Delta f},\quad \omega_{\mathrm{sfw0}}>>2\pi\,\Delta f, \label{eq:fw_group_c}
\end{align}
\end{subequations}

\noindent where $\omega_{\mathrm{efw0}}$, $\omega_{\mathrm{sfw}}$ represent the mechanical angular velocity of the flywheel rotor and the rated mechanical angular velocity, respectively. $J_{\mathrm{fw}}$, $J_{\mathrm{vir}}$ represent the physical inertia of FESS and the equivalent virtual inertia, respectively. The energy of FESS participating in the grid net load power regulation is given in \eqref{eq:fw_group_b}, further specified:  
\vspace{-0.2cm} 
\begin{subequations}\label{eq:Fess-const}
\begin{gather}
E^{\mathrm{F}}_{t} = E^{\mathrm{F}}_{t-1}-P_t^{\mathrm{F,s}}\Delta t
+P^{\mathrm{F,c}}_{t}\eta^{\mathrm{F,c}}\Delta t - \frac{P^{\mathrm{F,d}}_{t}}{\eta^{\mathrm{F,d}}}\Delta t, \label{eq:Fess-const_a}\\
P_t^{\mathrm{F,s}}=P^{\mathrm{wind}}(\omega_{\mathrm{sfw0}})+P^{\mathrm{brg}}(\omega_{\mathrm{sfw0}})+P_t^{\mathrm{bop}},\label{eq:Fess-const_b}\\
0 \le P^{\mathrm{F,c}}_{t} \le \overline{P}^{\mathrm{F}}\, \bigl(1-b^{\mathrm{F}}_{t}\bigr), \quad
0 \le P^{\mathrm{F,d}}_{t} \le \overline{P}^{\mathrm{F}}\, b^{\mathrm{F}}_{t}, \label{eq:Fess-const_c}\\
\underline{E}^{\mathrm{F}} \le E^{\mathrm{F}}_{t} \le \overline{E}^{\mathrm{F}}, \quad \forall\, t\in\mathbb{T},\label{eq:Fess-const_d}
\end{gather}
\end{subequations}

\noindent where $P_t^{\mathrm{F,s}}$ denotes the FESS standby loss. $P^{\mathrm{wind}}(\omega_{\mathrm{sfw0}})$, $P^{\mathrm{brg}}(\omega_{\mathrm{sfw0}})$, $P_t^{\mathrm{bop}}$ correspond to windage, bearing-friction, and auxiliary-consumption terms, respectively. Equation \eqref{eq:Fess-const} provides the iterative update for the FESS residual power and the associated operating constraints. $b^{\mathrm{F}}_{t}$ is a binary variable. Considering the degraded operation and maintenance costs of flywheel energy storage, its LCOE can be expressed as:
\vspace{-0.1cm}
\begin{subequations}
\label{eq:fess_lcos_cycle}
\begin{gather}
c_{\mathrm{lcoe}}^{\mathrm{F}}
=
\frac{
C_{\mathrm{inv}}^{\mathrm{F}}
+\displaystyle\sum_{n=1}^{\overline{N}^{\mathrm{F}}}\frac{c^{\mathrm{F}}_{\mathrm{om}}}{(1+r)^{n}}
}{
\displaystyle\sum_{n=1}^{\overline{N}^{\mathrm{F}}}\frac{\overline{E}^{\mathrm{F}}\,g(n)}{(1+r)^{n}}},\\
g(n)=1-n\beta^{\mathrm{F}},
\end{gather}
\end{subequations}

where $C^{\mathrm{F}}_{\mathrm{inv}}$ is the total investment of FESS, $c^{\mathrm{F}}_{\mathrm{om}}$ is the initial cycle operation and maintenance cost, $g(n)$ represents the capacity decay rate, and $\beta^{\mathrm{F}}$ is the linear decay factor.

\subsection{Dynamic Reconfigurable Battery Energy Storage Systems Modeling}\label{Battery Modeling}

\subsubsection{Structure of DRBN}

\begin{figure}[htbp] 
    \vspace{-0.2cm}
    \centerline{\includegraphics[width=1\columnwidth]{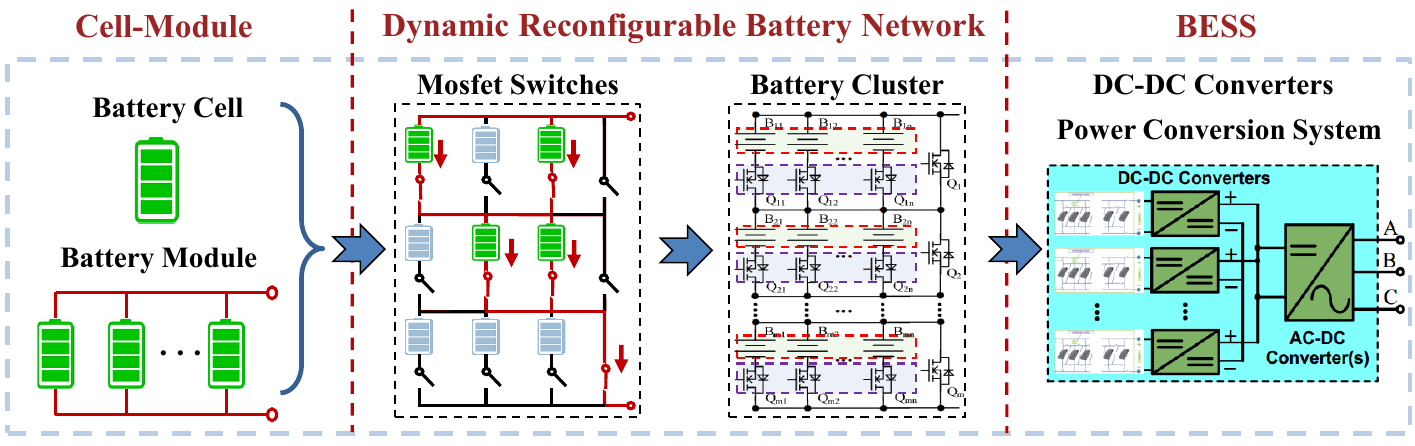}}
    \caption{DRBN system composition and topology.}
    \label{DRBN}
    \vspace{-0.3cm}
\end{figure}

Compared to traditional battery energy storage systems, the DRBN introduces a MOSFET switch array that enables second-level control for the battery modules involved in charging and discharging, as shown in Fig.~\ref{DRBN}. Specifically, for any series connection, real-time selection of high charge modules for participation in charging and discharging achieves synchronized charge variation across three parallel modules. The bypass switches are used to promptly isolate faulty battery modules or protect low-health-status batteries, ensuring synchronized health status changes within the system. As a result, the DRBN effectively avoids the "weak link effect" in terms of energy and lifespan within the system, extending the operating range of the storage system's full lifecycle from 80\% SOH to 20\% SOH, significantly reducing the levelized cost of electricity. Additionally, the BES includes DC/DC converters, Power Converter Systems, and other auxiliary devices.

\subsubsection{Real-Time Operating Cost Model for BESS}
A semi-empirical degradation model \cite{wan} is adopted to characterize the full-lifecycle degradation of the BESS.
\begin{equation}
\resizebox{\columnwidth}{!}{$
\displaystyle
\mathrm{SOH}
= \alpha_{\mathrm{sei}} e^{-\beta_{\mathrm{sei}} d_{\mathrm{sds}}}
+ \alpha_{\mathrm{sds}} e^{- d_{\mathrm{sds}}}
+ (1-\alpha_{\mathrm{sei}}-\alpha_{\mathrm{sds}})
  (1-\kappa e^{\beta_{\mathrm{cps}} d_{\mathrm{sds}}})
$,}
\end{equation}

\noindent where $\alpha_{\mathrm{sei}}$,$\alpha_{\mathrm{sds}}$ are the consumed capacity during SEI formation stage and steady degradation stage (SDS), $d_{\mathrm{sds}}$ degradation rate during SDS, $\beta_{\mathrm{sei}}$ and $\beta_{\mathrm{cps}}$ are proportional coefficients of the degradation rate of SEI formation stage and capacity knee stage. $d_{\mathrm{sds}}$ is determined by calendar aging and cyclic aging, and is affected by multiple stress factors and can be expressed as:
\vspace{-0.3cm}
\begin{subequations}\label{eq:damage_model}
\begin{align}
d_{\mathrm{sds}}(t,\tau,\nu,T)
&= d_t\!\bigl(t,\overline{\tau},\overline{T}\bigr)
 + \sum_{i=1}^{N} n_i\, d_c\!\bigl(\tau_i,\nu_i,T_i\bigr), \label{eq:damage_model_a}\\
d_t\!\bigl(t,\overline{\tau},\overline{T}\bigr)
&= S_t(t)\,S_{\tau}\!\bigl(\overline{\tau}\bigr)\,S_T\!\bigl(\overline{T}\bigr), \label{eq:damage_model_b}\\
d_c\!\bigl(\tau_i,\nu_i,T_i\bigr)
&= S_{\tau}(\tau_i)\,S_{\nu}(\nu_i)\,S_T(T_i), \label{eq:damage_model_c}
\end{align}
\end{subequations}

\noindent where $d_t$,$d_c$ represent calendar aging and cyclic aging rates, respectively. $S_t$, $S_{\tau}$, $S_T$, and $S_{\nu}$ are the stress factors for time, average SOC, average temperature, and DOD, respectively, can be expressed as:
\vspace{-0.2cm}
\begin{equation}
\begin{aligned}
&S_T(T) = \exp\!\left(k_T (T-T_{\mathrm{ref}})\,\frac{T_{\mathrm{ref}}}{273+T}\right), \quad
 S_t(t) = k_t\, t, \\
&S_{\tau}(\tau) = \exp\!\big(k_{\tau}(\tau-\tau_{\mathrm{ref}})\big), \quad
 S_{\nu}(\nu) = \frac{1}{\,k_{\nu1}\,\delta^{k_{\nu2}}+k_{\nu3}\,},
\end{aligned}
\label{eq:stress}
\end{equation}

\noindent where $k_T$, $k_t$, $k_{\tau}$ represent temperature, time, SOC stress factor, respectively. $k_{\nu1}$, $k_{\nu2}$, $k_{\nu3}$ are DOD stress factors.The specific values of all parameters are obtained by referring to \cite{wan}. Since all cells in the DRBN degrade at the same rate, the degradation of the BESS can likewise be described by the above formulas. The BESS's real-time operating cost is given by:

\vspace{-0.2cm}
\begin{equation}
    c_{\mathrm{cost}}^{\mathrm{B}}=\frac{C_{\mathrm{inv}}^{\mathrm{B}}}{80\%} \frac{\Delta L_c}{2E_{\mathrm{rated}}^{\mathrm{B}}}+c_{\mathrm{om}}^{\mathrm{B}},
\end{equation}

\noindent where $c_{\mathrm{cost}}^{\mathrm{B}}$, $c_{\mathrm{om}}^{\mathrm{B}}$ are BESS operating cost, OM cost per unit power per unit time, respectively. $C_{\mathrm{inv}}^{\mathrm{B}}$ represents BESS investment. $L_c=1-\mathrm{SOH}$, $\Delta L_c$ is the decay amount corresponding to one complete cycle.

\subsubsection{Operation constraints for BESS}For participation in power grid ancillary services, the BESS's state of chargeand dispatch power are constrained within the bounds specified below:
\vspace{-0.3cm}
\begin{subequations}\label{eq:bess-const}
\begin{gather}
E^{\mathrm{B}}_{t} = E^{\mathrm{B}}_{t-1}\!\left(1-\eta^{\mathrm{B,s}}\right)
+P^{\mathrm{B,c}}_{t}\eta^{\mathrm{B,c}}\Delta t - \frac{P^{\mathrm{B,d}}_{t}}{\eta^{\mathrm{B,d}}}\Delta t,
\quad \forall\, t\in\mathbb{T}\\
0 \le P^{\mathrm{B,c}}_{t} \le \overline{P}^{\mathrm{B}}\, \bigl(1-b^{\mathrm{B}}_{t}\bigr), \quad \forall\, t\in\mathbb{T}\\
0 \le P^{\mathrm{B,d}}_{t} \le \overline{P}^{\mathrm{B}}\, b^{\mathrm{B}}_{t}, \quad \forall\, t\in\mathbb{T}\\
\underline{E}^{\mathrm{B}} \le E^{\mathrm{B}}_{t} \le \overline{E}^{\mathrm{B}}, \quad \forall\, t\in\mathbb{T},
\end{gather}
\end{subequations}

\noindent where $\eta^{\mathrm{B,c}}$, $\eta^{\mathrm{B,d}}$, $\eta^{\mathrm{B,s}}$ represent charging efficiency, discharging efficiency, self-discharging rate, respectively. $b^{\mathrm{B}}_{t}$ is a binary variable.

\vspace{-0.5cm}
\subsection{Compressed Air Energy Storage Systems Modeling}\label{CAES Modeling}
\begin{figure}[htbp] 
    \vspace{-0.2cm}
    \centerline{\includegraphics[width=1\columnwidth]{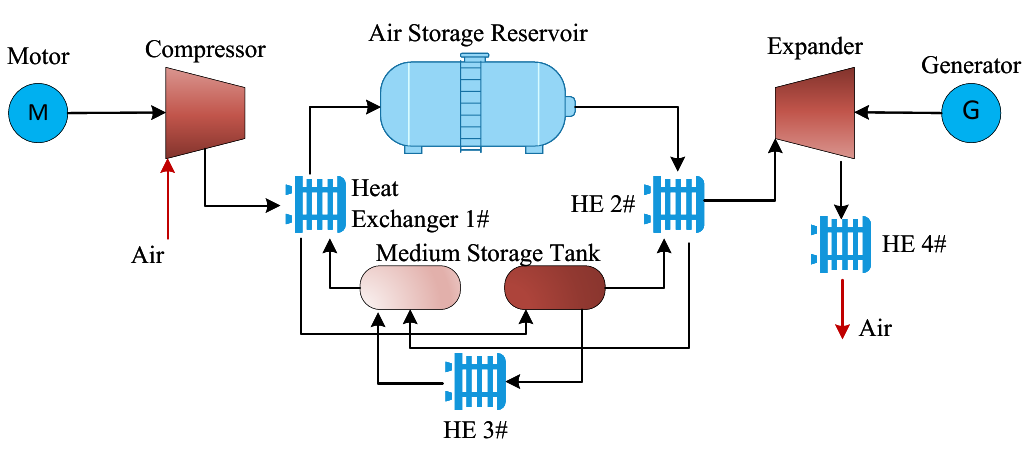}}
    \caption{CAES core components and operating logic.}
    \label{CAES}
    \vspace{-0.2cm}
\end{figure}

The CAES mainly consists of multi-stage compressor, thermal energy storage system (TES), air storage cavern, an expander,  heat exchanger, generator, and balance of plant (BOP), as shown in Fig.~\ref{CAES}. Electrical energy drives the multi-stage compressor via an electric motor, compressing the air while transferring the heat generated during compression to the TES for storage. During discharge, the stored thermal energy is transferred from the TES to the high-pressure air in the storage cavern through the heat exchanger, allowing the air to expand in the expander, which drives the generator to deliver electrical energy to the grid.

The compression and expansion of air rely on continuous thermodynamic processes, which limit the CAES's ability to provide second-level dynamic response, making it more suitable for minute to hour scale regulation. The dispatch model and operational constraints of CAES are similar to those in \eqref{eq:Fess-const_a}, \eqref{eq:Fess-const_c}, and \eqref{eq:Fess-const_d}, and the cost-per-energy model follows the same formulation as in \eqref{eq:fess_lcos_cycle}.
\vspace{-0.5cm}
\subsection{Hydrogen and Methanol Energy Storage Systems Modeling}\label{HM modeling}

\begin{figure}[htbp] 
    \vspace{-0.2cm}
    \centerline{\includegraphics[width=1\columnwidth]{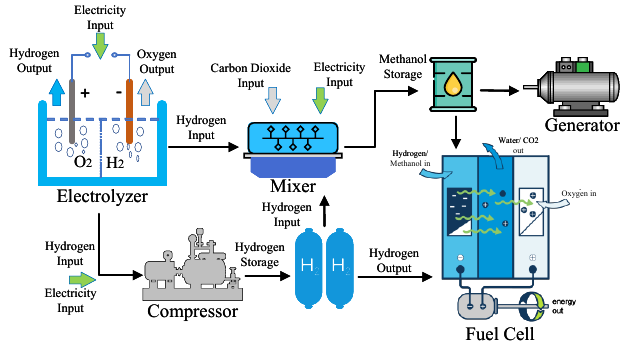}}
    \caption{Mechanism, core components and operating logic of HMES.}
    \label{HM}
    \vspace{-0.3cm}
\end{figure}

HMES consists of electrolyzer, compressor, hydrogen storage tank, mixer, fuel cell, generator, and other auxiliaries, as shown in Fig.~\ref{HM}. Due to the limited capacity of the hydrogen storage tank, hydrogen can be combined with carbon dioxide from industrial waste gases to form liquid methanol for large scale storage. Methanol can then be used for distributed power generation via fuel cells or for centralized power generation through combustion.

\subsubsection{Alkaline Water Electrolyzer Efiiciency}\label{HM charging modeling}
During AWE operation, direct current drives cathodic water reduction to generate hydrogen and anodic hydroxide oxidation to produce oxygen. The nonlinear electrochemical dynamics of the electrolyzer are characterized by the polarization curve, which is jointly governed by activation, ohmic, and mass transport overpotentials across distinct current regimes. Neglecting the mass transport effects, the terminal voltage of the AWE is formulated as:
\begin{equation} 
\begin{aligned} 
U_{\text{cell}}^{\mathrm{E}} &= U_{\text{rev}} + U_{\text{act,c}} + U_{\text{act,a}} + U_{\text{ohm}} = U_{\text{rev}} + \left[ (r_1 + d_1) + \right. \\
&\quad \left. r_2 \cdot \theta + d_2 \cdot P \right] \cdot \frac{i}{A} + s \cdot \log \left[ \left( t_1 + \frac{t_2}{\theta} + \frac{t_3}{\theta^2} \right) \cdot \frac{i}{A} + 1 \right], 
\end{aligned} 
\end{equation}

\noindent where $U_{\text{rev}}$, $U_{\text{act,c}}$, $U_{\text{act,a}}$, $U_{\text{ohm}}$ represent the reversible voltage, cathode activation over-potential, anode activation over-potential, and ohmic over-potential respectively. $i$ is the current of the electrolyzer, and $A$ is the effective area. The effects of temperature and pressure on the terminal voltage are reflected in $\theta$. The parameters $r_1$, $r_2$, $d_1$, $d_2$, $t_1$, $t_2$, $t_3$, and $s$ are constants that can be learned from experimental data \cite{H2_data}.

The charging efficiency of AWE is computed from the Faradaic efficiency $\eta_{\mathrm{F}}$, which is the ratio of the collected hydrogen molar rate to the theoretical rate implied by charge conservation. Major losses arise from hydrogen crossover from the cathode to the anode with re-oxidation, oxygen crossover to the cathode triggering oxygen reduction side reactions, and parasitic currents such as corrosion and shunt leakage. We adopt the following Faradaic-efficiency model:
\begin{equation}
\eta_{\mathrm{F}} = \frac{(i/A)^2}{\,f_{1} + f_{2}\,\cdot \theta + (i/A)^2\,}\,\bigl(f_{3} + f_{4}\,\cdot \theta\bigr),
\end{equation}

\noindent where parameters $f_{1},f_{2},f_{3},f_{4}$ are
the constants which can be learned from the experimental data. The hydrogen production rate and charging energy efficiency of AWE can be expressed as:
\vspace{-0.3cm}
\begin{subequations}
\begin{align}
h^{\mathrm{c}} &= 3600 \cdot \frac{\eta_{\mathrm{F}} \cdot M_{\mathrm{H}} \cdot i \cdot N}{2F},\\
\eta^{\mathrm{H,c}} &= \frac{h^{c} \cdot \mathrm{LHV_{\mathrm{H}}}}{P_{\text{stack}}}
= 3600 \cdot \frac{\eta_{\mathrm{F}} \cdot M_{\mathrm{H}} \cdot \mathrm{LHV_{\mathrm{H}}}}{2F \cdot U_{\text{cell}}},
\end{align}
\end{subequations}

\noindent where $h^{\mathrm{c}}$ represents AWE hydrogen production rate, $M_{\mathrm{H}}$ is the molar mass of hydrogen, i.e., $2.016\times10^{-3}~\mathrm{kg/mol}$, $F$ is the Faraday’s constant, i.e., $96485~\mathrm{C/mol}$, $N$ is the number of cells of the stack, and $\mathrm{LHV}_{\mathrm{H}}$ is the lower heat value of hydrogen, i.e., $33.33~\mathrm{kWh/kg}$.

\subsubsection{Hydrogen to Methanol Efficiency}\label{HM convert modeling}
The hydrogen to methanol conversion entails CO$_2$ capture, syngas conditioning, high-pressure catalytic synthesis, and subsequent product purification. For the energy efficiency evaluation, parasitic electrical loads (e.g., compressors and pumps) are omitted, as they account for merely 2-3\% of the methanol lower heating value (LHV). Similarly, external thermal demands are neglected, given that the required low to medium temperature heat is self-sustained through the exothermic synthesis and internal heat integration. Treating CO$_2$ as a zero energy carbon source, this analysis focuses exclusively on the chemical energy transition from hydrogen to methanol. The corresponding mass flow rates and LHV-based powers are formulated as:
\vspace{-0.2cm}
\begin{equation}
\begin{aligned}
&\dot m_{\mathrm{H}}^{\mathrm{in}} = \dot n_{\mathrm{H}}^{\mathrm{in}} M_{\mathrm{H}},
\dot m_{\mathrm{M}} = \dot n_{\mathrm{M}} M_{\mathrm{M}},
P^{\mathrm{H}} = \dot m_{\mathrm{H}}^{\mathrm{in}} \,\mathrm{LHV}_{\mathrm{H}}, \\
&P^{\mathrm{M}} = \dot m_{\mathrm{M}} \,\mathrm{LHV}_{\mathrm{M}}, 
X_{\mathrm{H}}= \frac{\dot n_{\mathrm{H}}^{\mathrm{in}} - \dot n_{\mathrm{H}}^{\mathrm{out}}}{\dot n_{\mathrm{H}}^{\mathrm{in}}},S_{\mathrm{H,M}}
= \frac{3\,\dot n_{\mathrm{M}}}{\dot n_{\mathrm{H}}^{\mathrm{in}} - \dot n_{\mathrm{H}}^{\mathrm{out}}},
\end{aligned}
\end{equation}

\noindent where $\dot n_{\mathrm{H}}^{\mathrm{in}}$, $\dot n_{\mathrm{H}}^{\mathrm{in}}$, $\dot n_{\mathrm{M}}$ are the hydrogen molar inlet rate, outlet, methonal molar flow rate, respectively. $P^{\mathrm{H}}$ and $P^{\mathrm{M}}$ denote the input power of hydrogen and the output power of methanol. $X_{\mathrm{H}}$, $S_{\mathrm{H,M}}$ represent hydrogen convertion ratio and selectivity to methanol, respectively. The methanol mass flow rate and mixer energy efficiency can be written as:
\vspace{-0.2cm}
\begin{subequations}
\begin{gather}
\dot m_{\mathrm{M}}
= \dot m_{\mathrm{H}}^{\mathrm{in}}\,
  X_{\mathrm{H}}\,
  S_{\mathrm{H,M}}\,\frac{M_{\mathrm{M}}}{3M_{\mathrm{H}}},\\
\eta^{\mathrm{H,M}}= \frac{P^{\mathrm{M}}}{P^{\mathrm{H}}}=X_{\mathrm{H}}\,S_{\mathrm{H,M}}\,\frac{M_{\mathrm{M}}}{3M_{\mathrm{H}}}\frac{\mathrm{LHV}_{\mathrm{M}}}{\mathrm{LHV}_{\mathrm{H}}}.
\end{gather}
\end{subequations}

To capture the dependence of efficiency on methanol output power, $X_{\mathrm{H}}$ and $S_{\mathrm{H,M}}$ are modeled as functions of the normalized load $\rho=P^{\mathrm{M}}/\overline{P}^{\mathrm{M}}$, $0 < \rho \le 1$. $\underline{\rho}$ represents the minimum stable load ratio, $\underline{X}_{\mathrm{H}}$ and $\overline{X}_{\mathrm{H}}$ are the conversions at $\rho=\underline{\rho}$ and $\rho=1$, respectively, and $\kappa_X>0$ a fitting parameter. To describe the decrease of conversion with increasing load, a saturation type model inspired by plug flow reactor behavior is adopted:
\vspace{-0.2cm}
\begin{equation}
\begin{aligned}
X_{\mathrm{H}}(\rho)
&= \underline{X}_{\mathrm{H}}
+ \bigl(\overline{X}_{\mathrm{H}}-\underline{X}_{\mathrm{H}}\bigr)\,
  \frac{1 - \exp(-\kappa_X/\rho)}
       {1 - \exp(-\kappa_X/\underline{\rho})},\\
S_{\mathrm{H,M}}(\rho)
&= \underline{S}_{\mathrm{H,M}}
+ \bigl(\overline{S}_{\mathrm{H,M}}-\underline{S}_{\mathrm{H,M}}\bigr)\,
  \exp\!\bigl[-\beta_S(\rho-\rho^\star)^2\bigr],
\end{aligned}
\end{equation}

\noindent where $\overline{S}_{\mathrm{H,M}}$ and $\underline{S}_{\mathrm{H,M}}$ are the maximum and minimum selectivity at an optimal load $\rho^\star$, respectively, and $\beta_S>0$ is a width parameter.

\subsubsection{HMES Discharging Effficiency}
PEMFC feature rapid startup capabilities and high power density. These characteristics enable swift responses to power demand fluctuations and seamless integration with renewable energy sources, making them highly suitable for the distributed generation scenarios based on hydrogen storage considered in this study. The terminal voltage of a single cell is governed by the reversible open circuit voltage and three primary voltage losses, which are formulated using an equivalent circuit model:
\vspace{-0.2cm}
\begin{subequations}
\begin{align}
    U_{\text{cell}}^{\mathrm{F}} &= E_{\text{Nernst}} - U_{\text{act}} - U_{\text{ohmic}} - U_{\text{con}}, \\
    E_{\text{Nernst}} &= \frac{1}{2F} \left[ \Delta G - \Delta S (\theta - \theta_{\text{ref}}) \right] \\
    &+ R \cdot \theta \left( \log(P_{\text{H}_2}) + \frac{\log(P_{\text{O}_2})}{2} \right), \notag \\
    U_{\text{act}} &= a_1 + a_2 \cdot \theta + a_3 \cdot \theta \log(C_{\text{O}_2}) + a_4 \cdot \theta \log(i), \\
    U_{\text{ohmic}} &= i \cdot R_{\text{ohmic}} = i \cdot \left( r_M \cdot l / A + R_c \right), \\
    U_{\text{con}} &= B \cdot \log \left( 1 - \frac{J}{J_{\text{max}}} \right), \quad J = \frac{i}{A},
\end{align}
\end{subequations}

\noindent where $E_{\text{Nernst}}$, $U_{\text{act}}$, $U_{\text{ohmic}}$, and $U_{\text{con}}$ denote the open circuit voltage alongside the activation, ohmic, and concentration overpotentials, respectively. Variables $\Delta G$ and $\Delta S$ represent the change in Gibbs free energy and entropy. $R$ is the universal gas constant (8.314 J/(K mol)), and $T_{\text{ref}}$ is the reference temperature (298.15 K). The terms $P_{\text{H}_2}$ and $P_{\text{O}_2}$ define the partial pressures of hydrogen and oxygen. $C_{\text{O}_2}$ represents the oxygen concentration at the cathode catalyst interface, $r_M$ is the electrolyte membrane resistivity, $l$ is the membrane thickness, and $B$ acts as the concentration overpotential coefficient. The operating and maximum current densities are given by $J$ and $J_{\text{max}}$, respectively. Finally, $a_1$, $a_2$, $a_3$, and $a_4$ are empirical constants derived from experimental data. The hydrogen consumption rate and discharging efficiency are formulated as:
\begin{subequations}
\begin{align}
    &h^{\mathrm{d}} = \frac{3600 \cdot M \cdot i \cdot N}{2F}, \\
    &\eta^{\mathrm{H,d}} = \frac{P_{\text{Stack}}}{h^{\mathrm{d}} \cdot HHV} = \frac{2F \cdot U_{\text{cell}}}{3600 \cdot M \cdot HHV},
\end{align}
\end{subequations}

\noindent where HHV is the higher heat value of hydrogen, i.e., 39.4 kWh/kg. For large-scale methanol-fired power generation via gas turbines, following standard modeling practices for gas turbine units, the methanol electrical output and methanol input power are related by a quadratic function. Accordingly, the methanol power generation efficiency can be expressed as:
\vspace{-0.2cm}
\begin{equation}
    \eta^{\text{M,d}}(x) = \overline{\eta}^{\text{M,d}}\bigl(1 - k(x - x^{\star})^{2}\bigr), 
    \qquad x = \frac{P}{P_{\text{rated}}},
\end{equation}
\noindent where $\overline{\eta}^{\text{M,d}}$ represent maximum efficiency of the unit at its optimal operating point, typically at 80\%-90\% of the rated load. $x^{\star}$ is relative load corresponding to the maximum efficiency, and $k$ is the curvature parameter of the efficiency curve.

\section{Full Timescale \\ Hierarchical MPC-MTIP Framework}\label{Method}

This section introduces the MPC structure of each layer in the full timescale hierarchical MPC-MTIP framework and the collaboration mode between layers. Before that, we introduce Motivation and outline.

\subsection{Motivation and outline}
\begin{figure}[htbp] 
    \vspace{-0.2cm}
    \centerline{\includegraphics[width=1\columnwidth]{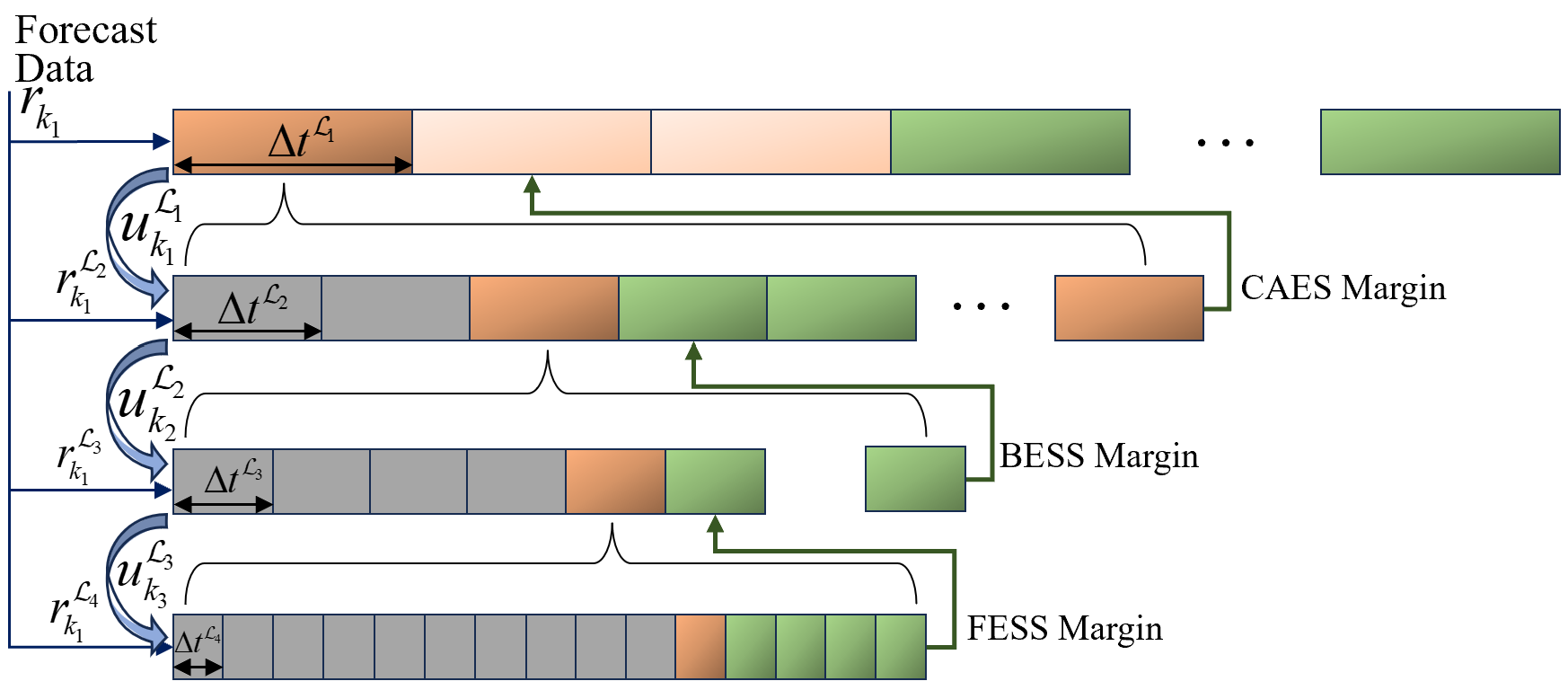}}
    \caption{Prediction, feedback, and decision flowchart of the hierarchical MPC-MTIP framework.}
    \label{flowchart}
    \vspace{-0.5cm}
\end{figure}

IMG is characterized by active power fluctuations at various timescales. At the same time, various energy storage systems exhibit multi-physics and cross-timescale heterogeneity. Using a single time step and a single controller to coordinate all energy storage systems can either force slower devices to chase faster fluctuations, resulting in poor efficiency and reduced lifespan, or shorten the forecast horizon, making inter-day/inter-week risks invisible. Moreover, periodic SOC constraints severely limit the flexibility of each storage unit in participating in net load smoothing. Based on this, this section proposes a control framework that considers the characteristics of various energy storage types, enabling flexible multi-layer coordination and balancing power supply security and economic efficiency, as shown in Fig.~\ref{flowchart}. 

Decisions regarding hydrogen storage power, methanol conversion rates, and methanol discharge are executed within the $\mathcal{L}_1$ layer. The $\mathcal{L}_2$, $\mathcal{L}_3$, and $\mathcal{L}_4$ layers consecutively optimize the intraday dispatch for the CAES, BESS, and FESS. Each hierarchical level exclusively mitigates the net load residual remains uncovered by the preceding tier and forwards the updated residual downward. Subordinate layers then supply compact margin signals upward, guiding the predictive controllers of higher tiers to determine the optimal dispatch for the subsequent receding horizon step. Through this downward propagation, the framework synchronously balances net load components from low to high frequencies. Concurrently, upward adjustments prevent capacity constrained subordinate devices from failing under sustained high frequency power shocks. Moreover, since methanol offers indefinite energy retention, it maximizes renewable utilization and secures the power supply against prolonged renewable generation deficits. The specific formulations for each predictive control layer are detailed in the remainder of this section.

\vspace{-0.2cm}
\subsection{Mathematical Model for Long-Duration Layer $\mathcal{L}_1$}

The $\mathcal{L}_1$ layer aims to address cross-seasonal power storage and load security by reducing the lowest frequency portion of the net load power curve. To account for factors influencing the long-term structure of net load, such as weather changes, while also ensuring accurate forecasting, a ten-day forecast horizon was chosen. To capture intraday structure, such as diurnal cycles, morning and evening peaks, and cloud coverage, a one-hour forecast granularity was chosen. The state space equations of $\mathcal{L}_1$ layer MPC can be written as:
\begin{subequations}\label{equation 8}
    \begin{align}
    \label{equation 8a}
     \boldsymbol{x}_{k_1+1}^{\mathcal{L}_1} &= \boldsymbol{A}\boldsymbol{x}_{k_1}^{\mathcal{L}_1} + \boldsymbol{B}^{\mathcal{L}_1} \boldsymbol{u}_{k_1}^{\mathcal{L}_1} \Delta t^{\mathcal{L}_1}, \quad \forall k_1 \in \mathbb{T}^{\mathcal{L}_1},\\
    \label{equation 8b}
     y_{k_1}^{\mathcal{L}_1} &= \boldsymbol{C}_y^{\mathcal{L}_1} \boldsymbol{u}_{k_1}^{\mathcal{L}_1},\\
    \label{equation 8c}
     \boldsymbol{A}=\boldsymbol{I}_2, 
     \boldsymbol{B}^{\mathcal{L}_1} &= \begin{bmatrix}
         \eta^{\mathrm{H,c}} & -1/\eta^{\mathrm{H,d}} & -1 & 0 \\
         0 & 0 & \eta^{\mathrm{M,c}} & -1/\eta^{\mathrm{M,d}}
     \end{bmatrix},\\
    \label{equation 8d}
      \boldsymbol{C}_y^{\mathcal{L}_1}&=[-1,1,0,1],
    \end{align}
\end{subequations}

\noindent where $\boldsymbol{x}_{k_1}^{\mathcal{L}_1}=[E_{k_1}^{\mathrm{H}},E_{k_1}^{\mathrm{M}}]^{\mathrm{T}}$, $\boldsymbol{u}_{k_1}^{\mathcal{L}_1}=[P_{k_1}^{\mathrm{H,c}},P_{k_1}^{\mathrm{H,d}},P_{k_1}^{\mathrm{M,c}},P_{k_1}^{\mathrm{M,d}}]^{\mathrm{T}}\geq 0$, represent the stored energy and power of hydrogen and methanol, respectively, $\Delta t^{\mathcal{L}_1}=1h$. Optimization model aims to minimize the net load power while reducing the instantaneous operating cost. The bi-objective optimization objective function of $\mathcal{L}_1$ layer can be formulated as follows:
\begin{equation}
\begin{split}
\min_{\boldsymbol{u}^{\mathcal{L}_1}} \Big[\, J(\boldsymbol{u}^{\mathcal{L}_1}, \boldsymbol{x}^{\mathcal{L}_1}) \triangleq
\sum_{i=k_1}^{k_1+N_1-1} \big( 
J_{\text{load}}^{\mathcal{L}_1}(i)+J_{\text{cost}}^{\mathcal{L}_1}(i)\big) \Big].
\end{split}
\label{eq:mpc_cost}
\end{equation}

\subsubsection{Grid Residual Net Load Penalty}

Denote by $\hat{P}_{\text{NL},k_1}^{\mathcal{L}_1}$ the hourly net load power forecast
and by $\hat{r}_{k_1}^{\mathcal{L}_1}$ the net load power to be saved at the $k_1$-th timestep in $\mathcal{L}_1$ layer. $\hat{P}_{\text{NL},k_1}^{\mathcal{L}_1}= \hat{r}_{k_1}^{\mathcal{L}_1}$. The unbalanced power after applying HMES can be expressed as:
\begin{equation}\label{marco}
\hat{P}_{\text{r},k_1}^{\mathcal{L}_1} = \hat{r}_{k_1}^{\mathcal{L}_1} - \boldsymbol{C}_y^{\mathcal{L}_1} \boldsymbol{u}_{k_1}^{\mathcal{L}_1}.
\end{equation}

The quadratic grid residual net load penalty term is formulated as:
\begin{equation}
J_{\text{load}}^{\mathcal{L}_1} = \sum_{i=k_1}^{k_1+N_1-1} (\hat{r}_i^{\mathcal{L}_1} - \boldsymbol{C}_y^{\mathcal{L}_1} \boldsymbol{u}_i^{\mathcal{L}_1})^\mathrm{T} Q^{\mathcal{L}_1} (\hat{r}_i^{\mathcal{L}_1} - \boldsymbol{C}_y^{\mathcal{L}_1} \boldsymbol{u}_i^{\mathcal{L}_1}),
\end{equation}

\noindent where, $Q^{\mathcal{L}_1}$ is the residual net load penalty weight coefficient. In order to ensure that $J_{\text{load}}^{\mathcal{L}_1}>0$, $Q^{\mathcal{L}_1}> 0, \forall i$.

\subsubsection{Linear Operational Cost}

The operational cost is calculated as a linear function of the control input power, multiplied by cost coefficients and the time step size $\Delta t^{\mathcal{L}_1}$:
\begin{equation}\label{J_cost}
J_{\text{cost}}^{\mathcal{L}_1} = \sum_{i=k_1}^{k_1+N_1-1} (\boldsymbol{c}^{\mathcal{L}_1})^\mathrm{T} R_i \boldsymbol{u}_i^{\mathcal{L}_1} \Delta t^{\mathcal{L}_1},
\end{equation}

\noindent where $R_i$ is the operation cost weight coefficient. The cost coefficient vector $\boldsymbol{c}^{\mathcal{L}_1}$ is defined as
$\boldsymbol{c}^{\mathcal{L}_1} = [c^{\mathrm{H,c}}, c^{\mathrm{H,d}}, c^{\mathrm{M,c}}, c^{\mathrm{M,d}}]^{\mathrm{T}}$.

\subsubsection{Complete Optimization Problem}

Combining the two terms above and related constraints, the complete optimization problem can be defined as:
\begin{equation}
   \begin{split}
\min_{\boldsymbol{u}^{\mathcal{L}_1}} \quad J^{\mathcal{L}_1}(\boldsymbol{u}^{\mathcal{L}_1}, \boldsymbol{x}^{\mathcal{L}_1}) = \sum_{i=k_1}^{k_1+N_1-1} \Big[(\hat{r}_i^{\mathcal{L}_1} - \boldsymbol{C}_y^{\mathcal{L}_1} \boldsymbol{u}_i^{\mathcal{L}_1})^\mathrm{T} Q_i \\ (\hat{r}_i^{\mathcal{L}_1} - \boldsymbol{C}_y^{\mathcal{L}_1} \boldsymbol{u}_i^{\mathcal{L}_1})+ (\boldsymbol{c}^{\mathcal{L}_1})^\mathrm{T} R_i \boldsymbol{u}_i^{\mathcal{L}_1} \Delta t^{\mathcal{L}_1}\Big],
   \end{split}
\end{equation}
\begin{subequations}\label{ur_constraints}
\begin{align}
    s.t. \quad &\underline{\boldsymbol{u}}_i^{\mathcal{L}_1} \leq \boldsymbol{u}_i^{\mathcal{L}_1} \leq \overline{\boldsymbol{u}}_i^{\mathcal{L}_1} \quad \forall i \in \mathbb{T}^{\mathcal{L}_1},\\
   & \underline{r}_{k_1}^{\mathcal{L}_1} \leq \hat{r}_{k_1}^{\mathcal{L}_1} - \boldsymbol{C}_y^{\mathcal{L}_1} \boldsymbol{u}_{k_1}^{\mathcal{L}_1} \leq \overline{r}_{k_1}^{\mathcal{L}_1}.
\end{align}
\end{subequations}

\noindent where $\underline{\boldsymbol{u}}_i^{\mathcal{L}_1}$, $\overline{\boldsymbol{u}}_i^{\mathcal{L}_1}$ are the upper and lower limits of energy storage power in $\mathcal{L}_1$ layer, $\underline{r}_{k_1}^{\mathcal{L}_1}$, $\overline{r}_{k_1}^{\mathcal{L}_1}$ represent the lower and upper bounds of the residual net load power at $k_1$-th time step, respectively, are determined by the underlying energy storage margin, which will be discussed in the next subsection. By stacking the control inputs, net load sequence, output variables, and operating-cost terms into vectors, the objective function can be succinctly expressed in the standard quadratic programming form as:
\begin{subequations}
\begin{gather}
J = (R_v - \mathcal{C}U)^\mathrm{T} (I_{N_1} \otimes Q)(R_v - \mathcal{C}U) + \mathcal{C}_{\text{cost}}^\mathrm{T} U, \\
\mathcal{C} = I_{N_1} \otimes C_y^{\mathcal{L}_1} \in \mathbb{R}^{N_1 \times 4N}, \,
\mathcal{C}_{\text{cost}} = (\mathbf{1}_{N_1} \otimes c^{\mathcal{L}_1})\Delta t \in \mathbb{R}^{4N_1 \times 1},\\
\boldsymbol{U} = \begin{bmatrix}\boldsymbol{u}_{k_1}^{\mathcal{L}_1} \\ \boldsymbol{u}_{k_1+1}^{\mathcal{L}_1} \\ \vdots \\ \boldsymbol{u}_{k_1+N_1-1}^{\mathcal{L}_1}\end{bmatrix} \in \mathbb{R}^{4N_1 \times 1}, \quad
\boldsymbol{R}_v = \begin{bmatrix}\hat{r}_{k_1} \\ \hat{r}_{k_1+1} \\ \vdots \\ \hat{r}_{k_1+N_1-1}\end{bmatrix} \in \mathbb{R}^{N_1 \times 1}.
\end{gather}
\end{subequations}

Expanding this expression, we obtain the standard form for quadratic programming, with matrices and vectors defined as:
\begin{subequations}
\begin{gather}
J = \frac{1}{2} U^\mathrm{T} H U + f^\mathrm{T} U + R_v^\mathrm{T}(I_{N_1} \otimes Q)R_v,\\
H = 2\mathcal{C}^\mathrm{T}(I_{N_1} \otimes Q)\mathcal{C}, \\
f = -2\mathcal{C}^\mathrm{T}(I_{N_1} \otimes Q)R_v + \mathcal{C}_{\text{cost}}.
\end{gather}
\end{subequations}

\subsection{Mathematical Model for $\mathcal{L}_2$, $\mathcal{L}_3$, $\mathcal{L}_4$ Layer}

CAES in $\mathcal{L}_2$ layer is scheduled to mitigate hour scale net load power fluctuations. The dominant uncertainties arise from wind turbine power affected by gusts, wind shear, and wind direction changes, PV output affected by variations in solar incidence angle and the passage of large-scale stratiform cloud systems, and load variations driven by industrial process cycles and equipment cycling. Accordingly, $\mathcal{L}_2$ adopts a 15min prediction and control granularity. The $\mathcal{L}_3$ BESS mitigates minute-scale net load fluctuations induced by turbulence, cloud motion, and therefore uses a 1min prediction and control resolution. The $\mathcal{L}_4$ FESS addresses second-scale power fluctuations via virtual inertia control. The quadratic grid residual net load penalty term of $\mathcal{L}_2$,$\mathcal{L}_3$ layer are expressed as:
\begin{subequations}
\begin{align}
J_{\text{load}}^{\mathcal{L}_2} 
&= \sum_{i=k_2}^{k_2+N_2-1} 
\bigl\|
\hat{r}_i^{\mathcal{L}_2} 
- \boldsymbol{C}_y^{\mathcal{L}_2} \boldsymbol{u}_i^{\mathcal{L}_2}
\bigr\|_{Q^{\mathcal{L}_2}}^{2},\\
J_{\text{load}}^{\mathcal{L}_3} 
&= \sum_{i=k_3}^{k_3+N_3-1} 
\bigl\|
\hat{r}_i^{\mathcal{L}_3} 
- \boldsymbol{C}_y^{\mathcal{L}_3} \boldsymbol{u}_i^{\mathcal{L}_3}
\bigr\|_{Q^{\mathcal{L}_3}}^{3},\\
\hat{r}_i^{\mathcal{L}_2}&=\hat{P}_{\text{NL},i}^{\mathcal{L}_2}-\boldsymbol{C}_y^{\mathcal{L}_1} \boldsymbol{u}_{k_1}^{\mathcal{L}_1},\\
\hat{r}_i^{\mathcal{L}_3}&=\hat{P}_{\text{NL},i}^{\mathcal{L}_3}- \boldsymbol{C}_y^{\mathcal{L}_1} \boldsymbol{u}_{k_1}^{\mathcal{L}_1}- \boldsymbol{C}_y^{\mathcal{L}_2} \boldsymbol{u}_{k_2}^{\mathcal{L}_2},
\end{align}
\end{subequations}

\noindent where $\hat{P}_{\text{NL},i}^{\mathcal{L}_2}$, $\hat{P}_{\text{NL},i}^{\mathcal{L}_3}$ represent the 15min and 1min granularity net load power forecast, respectively. In the hierarchical framework, one macro decision step $k_i$ of layer $\mathcal{L}_i$ corresponds to a specific set of micro steps in layer $\mathcal{L}_{i+1}$, denoted by $\mathbb{T}^{\mathcal{L}_{i+1}}$.$\boldsymbol{u}_{k_2}^{\mathcal{L}_2}=[P_{k_2}^{\mathrm{C,c}},P_{k_2}^{\mathrm{C,d}}]^{\rm T}\geq 0, \forall k_2 \in \mathbb{T}^{\mathcal{L}_2} $, $\boldsymbol{u}_{k_3}^{\mathcal{L}_3}=[P_{k_3}^{\mathrm{B,c}},P_{k_3}^{\mathrm{B,d}}]^{\rm T}\geq 0$, $\forall k_3 \in \mathbb{T}^{\mathcal{L}_3} $. $J_{\text{cost}}^{\mathcal{L}_2}$, $J_{\text{cost}}^{\mathcal{L}_3}$ take the same form as in \eqref{J_cost}, while $\boldsymbol{u}_{k_2}^{\mathcal{L}_2}, \boldsymbol{u}_{k_3}^{\mathcal{L}_3}$ are subject to the same constraints as in \eqref{ur_constraints}.

In $\mathcal{L}_4$ layer, explicit frequency dynamics are not modeled, instead, an equivalent frequency deviation is reconstructed from the residual net load imbalance, and a virtual inertia based control law is implemented using only power signals, which is written as:
\begin{equation}
r_{k_4}^{\mathcal{L}_4}=P_{\text{NL},k_4}- \boldsymbol{C}_y^{\mathcal{L}_1} \boldsymbol{u}_{k_1}^{\mathcal{L}_1}- \boldsymbol{C}_y^{\mathcal{L}_2} \boldsymbol{u}_{k_2}^{\mathcal{L}_2}-\boldsymbol{C}_y^{\mathcal{L}_3} \boldsymbol{u}_{k_3}^{\mathcal{L}_3}-P_{\text{NL},k_4}^\star,
\end{equation}

\noindent where $P_{\text{NL},k_4}$ denotes the instantaneous net load power, and $P_{\text{NL},k_4}^\star$ denotes the acceptable unbalanced power target trajectory. Under a standard aggregate static power-frequency characteristic, the equivalent frequency deviation $\Delta f_{k_4}$ satisfies:
\begin{equation}
  r^{\mathcal{L}_4}_{k_4} \approx K_{\text{pf}} \Delta f_{k_4},
  \label{eq:L4_r_f}
\end{equation}

\noindent where $K_{\text{pf}}>0$ denotes the aggregate power-frequency
characteristic coefficient of the system\cite{PFR}. In the discrete time implementation with sampling interval $\Delta t^{\mathcal{L}_4}$, the discrete derivative of the imbalance is approximated as:
\begin{equation}
  \dot{r}^{\mathcal{L}_4}_{i}
  \approx \frac{r^{\mathcal{L}_4}_{i} - r^{\mathcal{L}_4}_{i-1}}{\Delta t},
  \quad \forall i \in \mathbb{T}^{\mathcal{L}_4}.
  \label{eq:L4_r_dot}
\end{equation}

The virtual inertia based active power command of the FESS is given by:
\begin{equation}
  P_{i}^{F}
  = P_{0,i}
  - K_{\mathrm{P}}\, r^{\mathcal{L}_4}_{i}
  - K_{\mathrm{D}}\, \dot{r}^{\mathcal{L}_4}_{i},
  \quad \forall i \in \mathbb{T}^{\mathcal{L}_4},
  \label{eq:L4_Pref}
\end{equation}

\noindent where $K_{\mathrm{P}},K_{\mathrm{D}}>0$ are the effective droop and virtual inertia gains, respectively.

\subsection{MTIP-based Residual Net Load Bounds Determination}

To bridge the prediction granularity mismatch between the upper layer $\mathcal{L}_i$ and the lower layer $\mathcal{L}_{i+1}$, and to fully exploit the intrinsic temporal buffering potential of the lower layer ESS under fluctuating net loads, a feedback mechanism based on micro trajectory inverse projection (MTIP) is proposed. Unlike static constraints derived solely from instantaneous states, MTIP projects the high frequency dynamic capability of $\mathcal{L}_{i+1}$ into the decision space of the $\mathcal{L}_i$, rigorously ensuring solvability while optimizing cross layer economic synergy.

The macro residual net load $\hat{P}_{\text{r},k_i}^{\mathcal{L}_i}$, determined by the upper layer MPC decision variable $u_{k_i}^{\mathcal{L}_i}$, is given by \eqref{marco}. Conversely, the micro residual net load $\hat{r}_{k_{i+1}}^{\mathcal{L}_{i+1}}$ actually faced by the lower layer at any micro step $k_{i+1,j} \in \mathbb{T}^{\mathcal{L}_{i+1}}$ incorporates high granularity forecast information $\hat{P}_{\text{NL},k_{i+1}}^{\mathcal{L}_{i+1}}$. To explicitly model the cross-scale coupling, we decompose the micro residual into the controllable macro baseline and an uncontrollable micro differential fluctuation $\boldsymbol{\xi}_{k_{i+1}}$:
\begin{equation}
\hat{r}_{k_{i+1,j}}^{\mathcal{L}_{i+1}} = \hat{P}_{\text{r},k_i}^{\mathcal{L}_i} + \xi_{k_{i+1,j}}, \quad \forall k_{i+1,j} \in \mathbb{T}^{\mathcal{L}_{i+1}},
\label{eq:micro_decomposition}
\end{equation}

\noindent where $\xi_{k_{i+1,j}} \triangleq \hat{P}_{\text{NL},k_{i+1}}^{\mathcal{L}_{i+1}} - \hat{r}_{k_i}^{\mathcal{L}_i}$ represents the high frequency power component that is invisible to the upper layer $\mathcal{L}_i$ but mandatory for the lower layer to absorb. A constant macro baseline $\hat{P}_{\text{r},k_i}^{\mathcal{L}_i}$ is physically admissible if and only if the resulting state trajectory of the lower-layer ESS remains within the safety envelope $(\underline{E}^{\mathcal{L}_{i+1}}, \overline{E}^{\mathcal{L}_{i+1}})$ for all micro steps $k_{i+1},j \in \mathbb{T}^{\mathcal{L}_{i+1}}$. The corresponding ESS's energy state evolution at step $k_{i+1},j$ is given by.:
\begin{equation}
E_{k_{i+1},j}^{\mathcal{L}_{i+1}} = E_{k_{i+1},0}^{\mathcal{L}_{i+1}} - \sum_{m=1}^{j} \Psi\left( \hat{P}_{\text{r},k_i}^{\mathcal{L}_i} + \xi_{k_{i+1},m} \right) \Delta t^{\mathcal{L}_{i+1}},
\label{eq:energy_dynamics}
\end{equation}

\noindent where $\Psi(\cdot)$ characterizes the corrected interaction power between the ESS and the grid, considering charging efficiency $\eta_{\mathrm{c}}^{\mathcal{L}_{i+1}}$ and discharging efficiency $\eta_{\mathrm{d}}^{\mathcal{L}_{i+1}}$ of layer $\mathcal{L}_{i+1}$:
\begin{equation}\label{psi}
\Psi(P) = \begin{cases} P / \eta_{\mathrm{d}}^{\mathcal{L}_{i+1}}, & P \ge 0 \\ P \cdot \eta_{\mathrm{c}}^{\mathcal{L}_{i+1}}, & P < 0 \end{cases}
\end{equation}

The rigorous physical feasible region $(\underline{r}_{\text{phy}, k_i}^{\mathcal{L}_i}, \overline{r}_{\text{phy}, k_i}^{\mathcal{L}_i})$ for $\hat{P}_{r,k_1}^{\mathcal{L}_1}$ can be derived based on \eqref{psi}. To rigorously quantify the energy buffering contribution of the fluctuations, we introduce the integrated forecast residual energy, $\Phi_{k_{i+1},j}^{+}(\boldsymbol{\xi}_{k_{i+1}}),\Phi_{k_{i+1},j}^{-}(\boldsymbol{\xi}_{k_{i+1}})$. To ensure dimensional consistency, power fluctuations are explicitly integrated over the micro intervals:
\begin{subequations}
\begin{align}
\Phi_{k_{i+1},j}^{+}(\boldsymbol{\xi}_{k_{i+1}}) &= \sum_{m=1}^{j} \left[ \phi^{+}\left( \xi_{k_{i+1},m} \right) \cdot \Delta t^{\mathcal{L}_{i+1}} \right],\\
\Phi_{k_{i+1},j}^{-}(\boldsymbol{\xi}_{k_{i+1}}) &= \sum_{m=1}^{j}\left[ \phi^{-}\left( \xi_{k_{i+1},m} \right) \cdot \Delta t^{\mathcal{L}_{i+1}} \right],
\label{eq:fluctuation_potential}
\end{align}
\end{subequations}

\noindent where $\phi^{+}(x) = x$ for $x \ge 0$, and $\phi^{+}(x) = x \cdot (\eta_{\mathrm{d}}^{\mathcal{L}_{i+1}}\eta_{\mathrm{d}}^{\mathcal{L}_{i+1}})$ for $x < 0$. $\phi^{-}(x) = x/(\eta_{\mathrm{c}}^{\mathcal{L}_{i+1}}\eta_{\mathrm{d}}^{\mathcal{L}_{i+1}})$ for $x \ge 0$, and $\phi^{+}(x) = x$ for $x < 0$. The physical feasible region $(\underline{r}_{\text{phy}, k_i}^{\mathcal{L}_i}, \overline{r}_{\text{phy}, k_i}^{\mathcal{L}_i})$ is constrained by the energy trajectory's extrema and the safety envelope, $(\underline{E}^{\mathcal{L}_{i+1}}, \overline{E}^{\mathcal{L}_{i+1}})$ of ESS in layer $\mathcal{L}_{i+1}$, which can be expressed as:
\begin{subequations}
\begin{align}
\overline{r}_{\text{phy}, k_i}^{\mathcal{L}_i} &= \min_{\mathbb{T}_{k_i}^{\mathcal{L}_{i+1}}}  \frac{ (E_{k_{i+1},j}^{\mathcal{L}_{i+1}} - \underline{E}^{\mathcal{L}_{i+1}})\eta_d^{\mathcal{L}_{i+1}} - \Phi_{k_{i+1},j}^{+}(\boldsymbol{\xi}_{k_{i+1}}) }{ j \cdot \Delta t^{\mathcal{L}_{i+1}} } ,\\
\underline{r}_{\text{phy}, k_i}^{\mathcal{L}_i} &= \max_{\mathbb{T}_{k_i}^{\mathcal{L}_{i+1}}}  \frac{ (E_{k_{i+1},j}^{\mathcal{L}_{i+1}} - \overline{E}^{\mathcal{L}_{i+1}})/\eta_c^{\mathcal{L}_{i+1}} - \Phi_{k_{i+1},j}^{-}(\boldsymbol{\xi}_{k_{i+1}}) }{j\cdot \Delta t^{\mathcal{L}_{i+1}} }.
\label{eq:upper_lower_bound_phy}
\end{align}
\end{subequations}

The term $-\Phi_{k_{i+1}}^{+}(\boldsymbol{\xi}),-\Phi_{k_{i+1}}^{-}(\boldsymbol{\xi})$ acts as virtual capacity for lower layer $\mathcal{L}_{i+1}$ , mathematically validating that $\overline{r}_{\text{phy}, k_i}^{\mathcal{L}_i}$ can be positive even when the initial energy is near minimum, $E_{k_{i+1},0}^{\mathcal{L}_{i+1}} \to \underline{E}^{\mathcal{L}_{i+1}}$, and $\underline{r}_{\text{phy}, k_i}^{\mathcal{L}_i}$ can be negative even when the initial energy is near maximum, $E_{k_{i+1},0}^{\mathcal{L}_{i+1}} \to \overline{E}^{\mathcal{L}_{i+1}}$. This mechanism enables the lower layer ESS to leverage high frequency fluctuations to facilitate upper layer energy throughput, even under conditions of low available margins.

While the physical domain $[\underline{r}_{\text{phy}, k_i}^{\mathcal{L}_i}, \overline{r}_{\text{phy}, k_i}^{\mathcal{L}_i}]$ ensures feasibility, economic optimality is governed by the relative marginal costs. Let $\lambda^{\mathcal{L}_i}$ and and $\lambda^{\mathcal{L}_{i+1}}$be the marginal cost of the upper and lower layer, which can be derived by:
\begin{equation}
\lambda^{\mathcal{L}_i} =\frac{\partial J_{\text{cost}}^{\mathcal{L}_i}}{\partial (u^{\mathcal{L}_i} \Delta t^{\mathcal{L}_i})}= R_{k_i}^T c^{\mathcal{L}_i},
\end{equation}

The final adaptive bounds $(\underline{r}_{k_i}^{\mathcal{L}_i}, \overline{r}_{k_i}^{\mathcal{L}_i})$ fed back to layer $\mathcal{L}_i$ can be derived as:
\begin{subequations}
\begin{align}
\overline{r}_{k_i}^{\mathcal{L}_i} &= \Gamma_{k_i} \cdot \overline{r}_{\text{phy}, k_i}^{\mathcal{L}_i} + (1 - \Gamma_{k_i}) \cdot \epsilon_{\text{db}}, \\
\underline{r}_{k_i}^{\mathcal{L}_i} &= \Gamma_{k_i} \cdot \underline{r}_{\text{phy}, k_i}^{\mathcal{L}_i} - (1 - \Gamma_{k_i}) \cdot \epsilon_{\text{db}},\\
\Gamma_{k_i} &= \left[ 1 + \exp\left( -\kappa \cdot \left( \frac{\lambda^{\mathcal{L}_i}}{\lambda^{\mathcal{L}_{i+1}}} - \chi_{\text{th}} \right) \right) \right]^{-1},
\label{eq:elastic_factor}
\end{align}
\label{eq:final_bounds}
\end{subequations}

\noindent where $\Gamma_{k_i} \in (0, 1]$ is the weighting coefficient for the adaptive bounds, accounting for the marginal costs of the upper and lower layers.$\epsilon_{\text{db}}$ represents a minimal deadband. $\kappa$, $\chi_{\text{th}}$ denote the constant slope and threshold of the sigmoid function, respectively.

\section{Results and Discussions}\label{Experiments}

\subsection{Experimental Setups and Design}\label{set up}

\begin{figure}[htbp] 
    \vspace{-0.1cm}
    \centerline{\includegraphics[width=1\columnwidth]{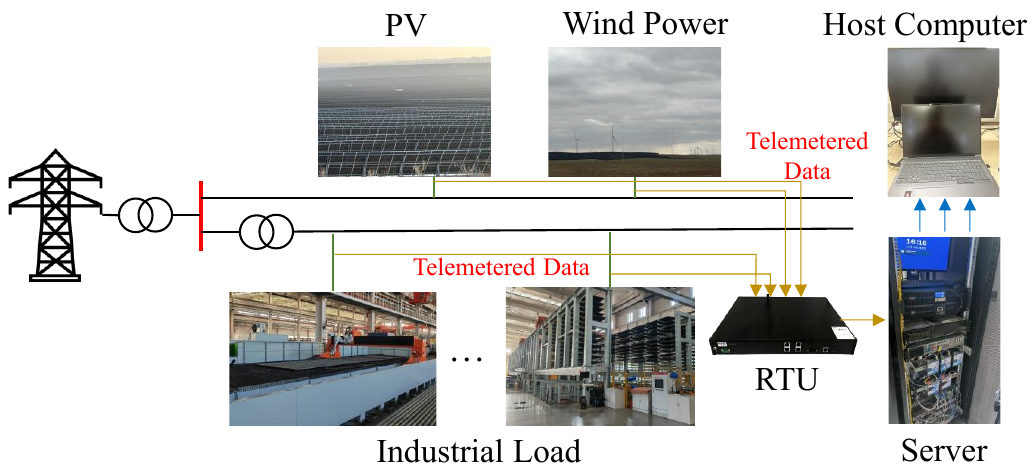}}
    \caption{Distribution and variation trends of wind and photovoltaic power output over a 14-month period.}
    \label{experiment}
    \vspace{-0.1cm}
\end{figure}

\begin{figure}[htbp] 
    \vspace{-0.2cm}
    \centerline{\includegraphics[width=1\columnwidth]{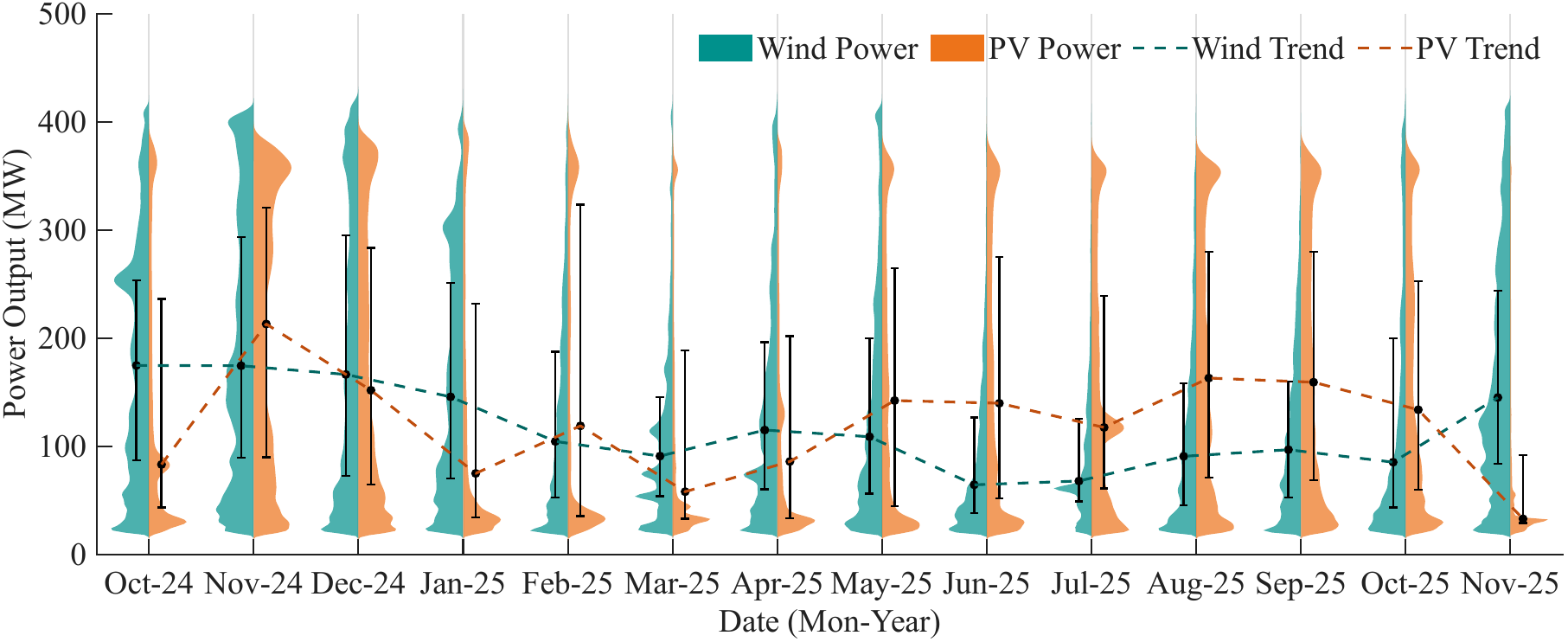}}
    \caption{Distribution and variation trends of wind and photovoltaic power output over a 14-month period.}
    \label{violin}
    \vspace{-0.2cm}
\end{figure}

\begin{figure}[htbp] 
    \vspace{-0.2cm}
    \centerline{\includegraphics[width=1\columnwidth]{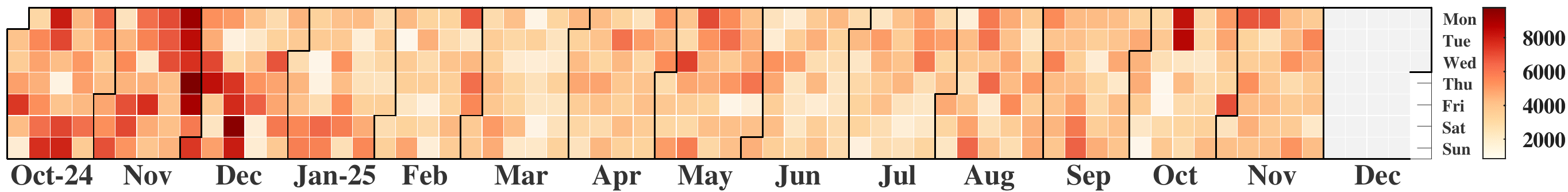}}
    \caption{Statistics on daily energy generation of renewable energy units.}
    \label{heatmap}
    \vspace{-0.2cm}
\end{figure}

The empirical validation employs a 14-month dataset acquired from a renewable energy base and a neighboring industrial park in Inner Mongolia, China, as shown in Fig.~\ref{experiment}. The site features a wind capacity of $425\text{MW}$ and a PV capacity of $375\text{MW}$, serving a predominantly industrial and commercial load of $200\text{MW}$. RES Data were sampled at 1s interval. Fig.~\ref{violin} presents the long-term power output profiles and their monthly distributions, and Fig.~\ref{heatmap} displays the total daily energy generation. 


\vspace{-0.5cm}
\begin{table}[htbp]
\caption{Capacity, Power Configurations and Parameter Settings of the HESS}
\label{configuration}
\centering
% 将列间距从默认的 6pt 缩小到 3pt
\setlength{\tabcolsep}{2pt} 
% 在 c, l, l 前后加上 @{} 去除表格边缘的空白
\begin{tabular}{@{} c l l @{}}
\toprule
\textbf{Layer} & \textbf{Symbol} & \textbf{Value \& Unit} \\
\midrule
$\mathcal{L}_1$ 
& $\overline{P}^H, \overline{E}^H$ & 200 MW, 5000 MWh \\
& $\overline{P}^M, \overline{E}^M$ & 200 MW, $\infty$ \\
& $\overline{\eta}^{H,c}, \overline{\eta}^{H,d}, \overline{\eta}^{H,M},\overline{\eta}^{M,d}$ & 0.65-0.85, 0.62-0.78, 0.7-0.8, 0.52 \\
\midrule
$\mathcal{L}_2$ 
& $\overline{P}^C, \overline{E}^C$ & 100 MW, 1000 MWh \\
& $\eta^{C,c}, \eta^{C,d}$ & 0.83, 0.83 \\
\midrule
$\mathcal{L}_3$ 
& $\overline{P}^B, \overline{E}^B$ & 100 MW, 200 MWh \\
& $\eta^{B,c}, \eta^{B,d}$ & 0.95, 0.95 \\
\midrule
$\mathcal{L}_4$ 
& $\overline{P}^F, \overline{E}^F$ & 60 MW, 2 MWh \\
& $\eta^{F,c}, \eta^{F,d}$ & 0.95, 0.95 \\
& $K_{pf}, K_P$ & 15 MW/Hz, 120 MW/Hz\\
& $K_D$ &25 MW$\cdot$s/Hz \\
\midrule
MPC  
& $Q^{\mathcal{L}_1}/R^{\mathcal{L}_1}, Q^{\mathcal{L}_2}/R^{\mathcal{L}_2}, Q^{\mathcal{L}_3}/R^{\mathcal{L}_3}$ & 0.27, 0.15, 0.1 \\
MTIP & $\chi_{th}, \kappa, \epsilon_{db}$ & 1.0, 5.0, 2 MW \\
\bottomrule
\end{tabular}
\vspace{-0.7cm}
\end{table}

\subsection{Effectiveness and Robustness Test of the proposed method under Different Scenarios and Prediction Accuracies}\label{Test}

\begin{figure}[htbp] 
    \vspace{-0.2cm}
    \centerline{\includegraphics[width=1\columnwidth]{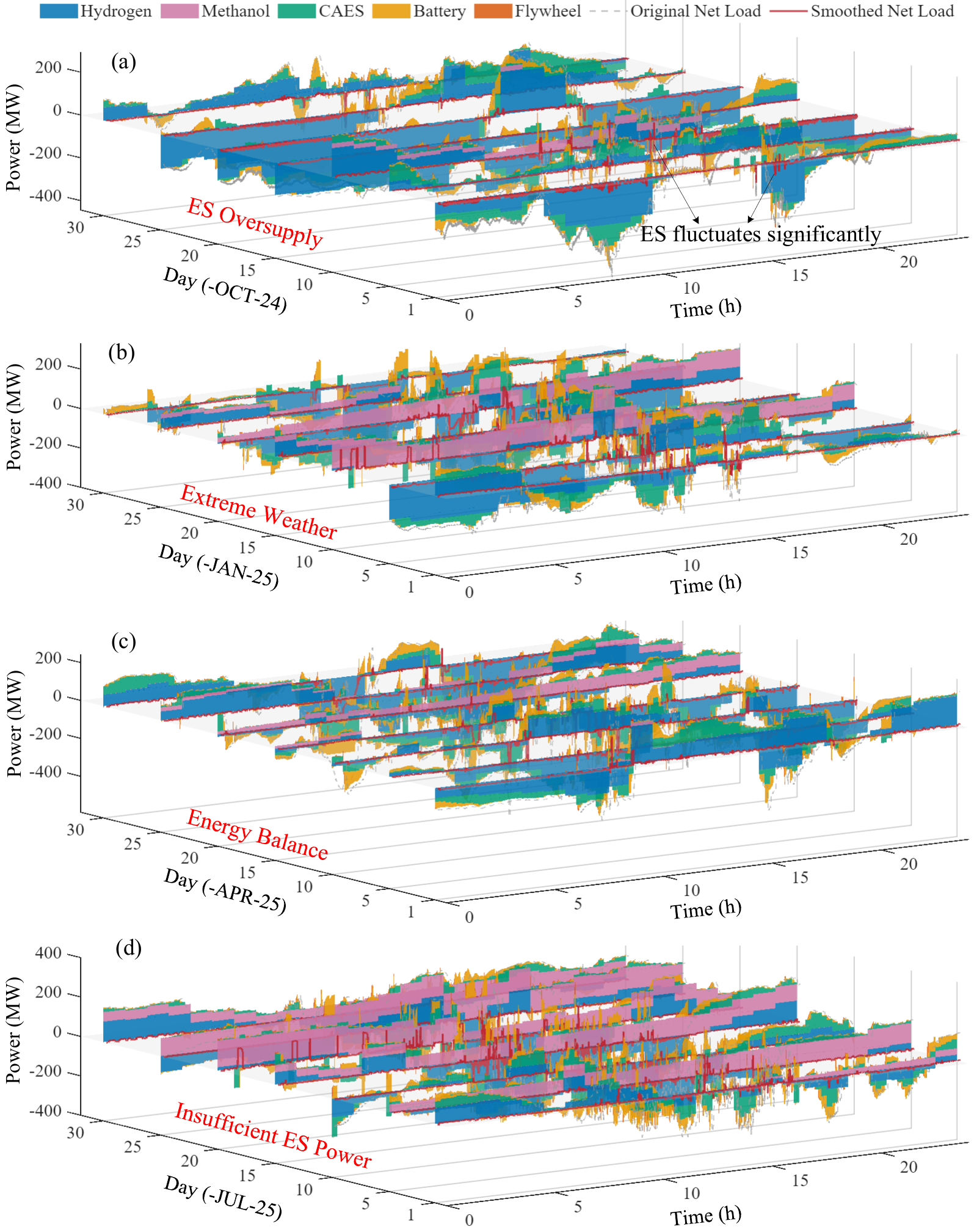}}
    \caption{HESS output and net load variations under different scenarios: (a) ES oversupply month, (b) extreme Weather month, (c) energy balance month and (d) insufficient ES month.}
    \label{Manage}
    \vspace{-0.3cm}
\end{figure}

To demonstrate its effectiveness and robustness, the proposed method is validated using continuous 14-month data, and analyzed under several typical scenarios. The capacity, power configurations, and parameter settings of the HESS are summarized in Table~\ref{configuration}. Fig. \ref{Manage} illustrates the NL smoothing performance of the proposed method under a 90\% forecast accuracy across four seasonal supply-demand profiles and extreme scenarios. In autumn (Fig. \ref{Manage}(a)), abundant renewables and light loads allow $\mathcal{L}_1$ to primarily charge, storing excess hydrogen as methanol. Notably, the fixed decision steps across $\mathcal{L}_1$--$\mathcal{L}_3$ leave residual spikes when instantaneous NL fluctuations exceed the flywheel's rated power. Conversely, during the extreme zero-renewable period (Jan. 10--20, Fig. \ref{Manage}(b)), $\mathcal{L}_1$ pre-charges based on forecasts. Storage tiers are dispatched at high frequencies, with batteries and CAES rapidly cycling atop the hydrogen baseload. This validates the hierarchical dynamic response and demonstrates that the MTIP feedback mechanism effectively prevents SOC limit induced deadlocks in lower-tier storages under severe fluctuations. Under balanced spring conditions (Fig. \ref{Manage}(c)), the operational cost term in the MPC objective function avoids unnecessary storage dispatch during minor fluctuations. During summer (Fig. \ref{Manage}(d)), prolonged power deficits caused by weak wind and high industrial loads are mitigated by continuous high power methanol integration, ensuring supply security. Over the 14-month operational period, the proposed method enables the HESS to achieve average reduction rates of 97.4\% for the net load and 92.7\% for minute-level power fluctuations, while maintaining an average round-trip efficiency of 62.2\%.
\begin{figure}[htbp] 
    \vspace{-0.2cm}
    \centerline{\includegraphics[width=1\columnwidth]{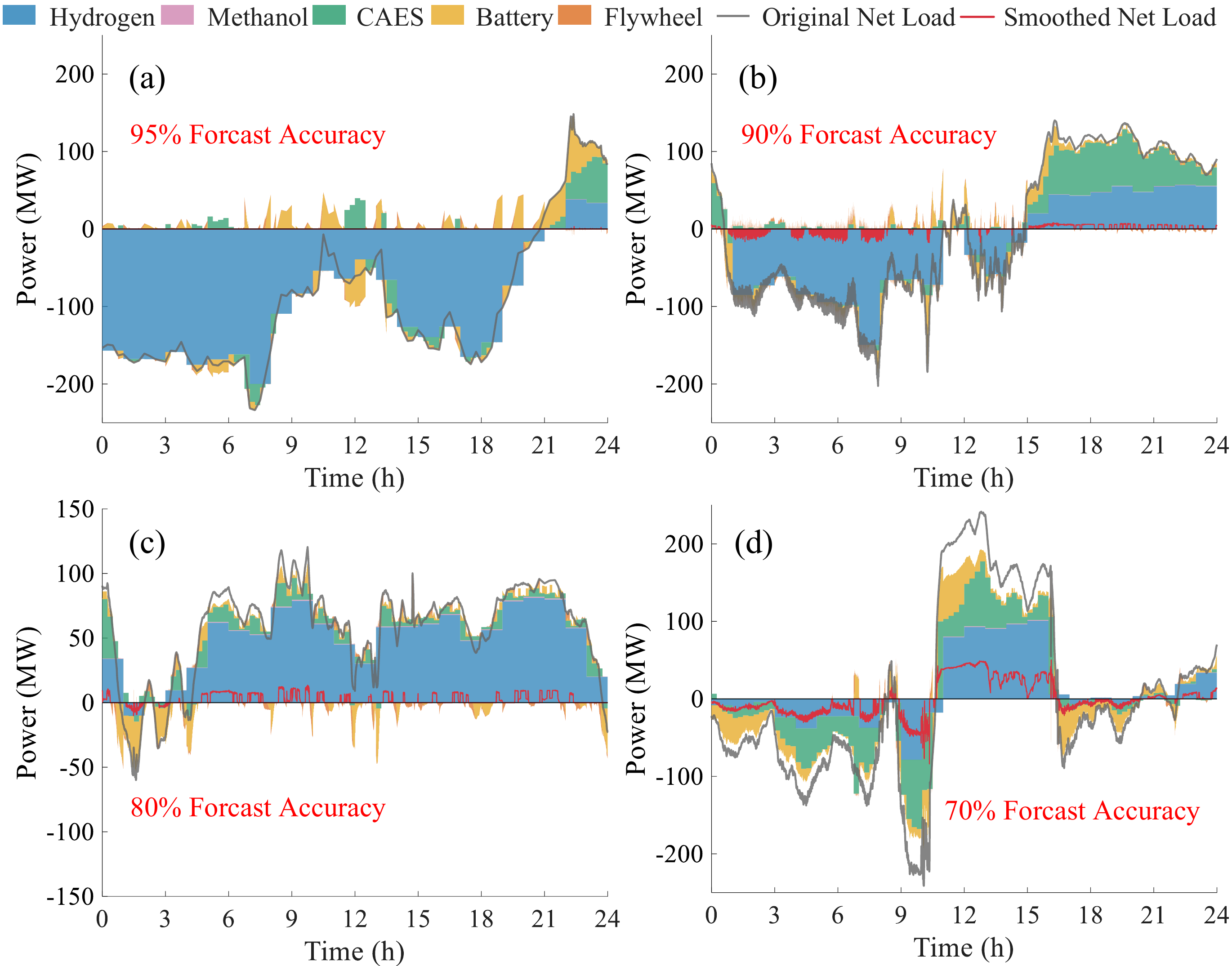}}
    \caption{Robustness of the proposed method under varying forecast accuracies: (a) 95\% (Oct. 15, 2024), (b) 90\% (Oct. 16, 2024), (c) 80\% (Oct. 17, 2024) and (d) 70\% (Oct. 18, 2024).}
    \label{robust}
    \vspace{-0.6cm}
\end{figure}

\subsection{Comparison of the Effects and Economic Analysis of Different Methods}\label{Compare}
Fig. \ref{robust} demonstrates the robustness of the proposed method under varying net load forecast accuracies. At a 95\% accuracy, 99.6\% of the net load is successfully smoothed. Performance remains consistent across the 80\%--90\% accuracy range, as the bottom-tier flywheel, employing virtual inertia control, partially mitigates the net load residuals left by the multi-layer MPC. At a 70\% accuracy, these residuals significantly exceed the flywheel's rated capacity, causing a performance drop; nevertheless, smoothing 83.2\% of the net load still evidences the method's strong robustness.

\begin{figure}[htbp] 
    \vspace{-0.2cm}
    \centerline{\includegraphics[width=1\columnwidth]{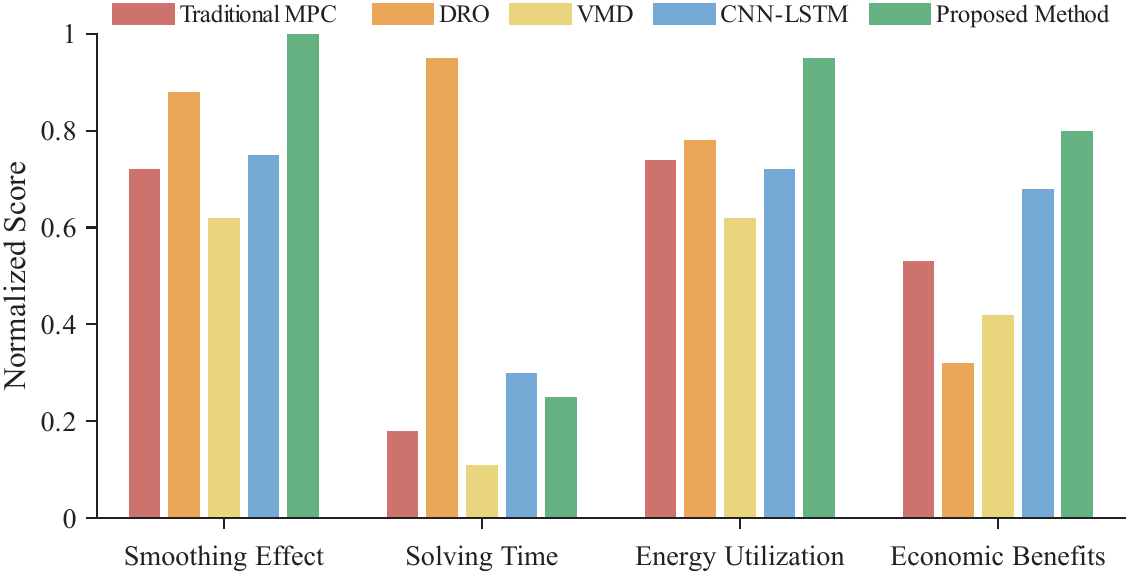}}
    \caption{Performance comparison of different methods.}
    \label{compare}
    \vspace{-0.1cm}
\end{figure}

Fig. \ref{compare} compares the multi-indicator performance of the proposed method against mainstream HESS energy management models. Specifically, the Traditional MPC employs an existing model with periodic SOC constraints, while the CNN-LSTM is trained using 13-month data alongside \textit{a posteriori} optimal storage dispatch results. The proposed method exhibits significant advantages in net load smoothing, energy utilization, and economic benefits. However, owing to the MTIP feedback mechanism, its computational efficiency is marginally lower than that of frequency-domain decomposition and Traditional MPC methods. Nevertheless, on an Intel Core Ultra 9 275HX @ 2.70 GHz with 32 GB RAM, the solving time per hour (3600 steps) remains under 0.02 s. The economic benefits are calculated by considering the electricity savings of the industrial microgrid under a two-part tariff and the operational costs of the HESS, detailed as:
\vspace{-0.3cm}
\begin{equation}
   \begin{split}
B = &\sum_{t=1}^{T} c_{tou}(t) \left( P_{L}(t) - P_{g}(t) \right) \Delta t 
\\&+ c_{cap} \left( \max_{t \in T} P_{L}(t) - \max_{t \in T} P_{g}(t) \right) - \sum_{i=1}^{4}J_{\text{cost}}^{\mathcal{L}_i},
   \end{split}
\end{equation}
\noindent where $c_{tou}(t)$ and $c_{cap}$ denote the time-of-use electricity price and the capacity charge rate, respectively. $P_{L}(t)$ and $P_{g}(t)$ are the original and smoothed  net load.

\section{Conclusion}\label{CONCLUSION}
In this paper, a full-timescale hybrid energy storage model tailored for IMG is established and a novel hierarchical MPC framework to address energy management challenges is proposed. Discarding traditional periodic SOC hard constraints, we introduced an adaptive feedback mechanism based on MTIP to accurately quantify the dynamic margins of lower layer storage units during net load smoothing. Validated using 14 consecutive months of second-level data from a real-world IMG, experimental results demonstrate that the proposed method achieves an impressive average net load smoothing rate of 97.4\% and a minute-level fluctuation elimination rate of 92.7\%, while maintaining a comprehensive cycle efficiency of 62.2\%. Furthermore, the framework exhibits strong robustness by sustaining an 83.2\% smoothing rate even under a low prediction accuracy of 70\%. Overall, it significantly outperforms existing mainstream methods across multiple dimensions, including smoothing efficacy, energy utilization, and economic benefits.

Future work will focus on the optimal capacity and power sizing for individual energy storage units. Additionally, further research will integrate resilient response strategies into the MPC-MTIP framework to address exceptional scenarios, such as specific storage failures or emergency grid dispatch events.

\bibliographystyle{IEEEtran}
\bibliography{HESS}

\end{document}